\documentstyle[12pt]{article}
\global\parskip 6pt
\begin{document}
\begin{titlepage}
\begin{center}
{\Large\bf 1992 Trieste Lectures on}\\
\vskip .1in
{\Large\bf Topological Gauge Theory}\\
\vskip .1in
{\Large\bf And Yang-Mills Theory}\\
\vskip .5in
{\bf George Thompson} \footnote{e-mail:
                      thompson@itsictp.it}
                     \\
\vskip .10in
{\em I.C.T.P.} \\
{\em P.O. BOX 586}\\
{\em 34100 TRIESTE, ITALY}\\
\end{center}
\vskip .10in
\begin{abstract}

\end{abstract}
\end{titlepage}

\newcommand{\be}{\begin{equation}}
\newcommand{\ee}{\end{equation}}
\newcommand{\bea}{\begin{eqnarray}}
\newcommand{\eea}{\end{eqnarray}}
\newcommand{\p}{\psi}
\newcommand{\f}{\phi}
\newcommand{\fb}{\bar{\phi}}
\newcommand{\vf}{\varphi}
\renewcommand{\d}{\delta}
\newcommand{\e}{\eta}
\newcommand{\ep}{\epsilon}
\renewcommand{\c}{\chi}
\renewcommand{\l}{\lambda}
\newcommand{\m}{\mu}
\newcommand{\n}{\nu}
\renewcommand{\a}{\alpha}
\newcommand{\A}{{\cal A}}                 
\newcommand{\AG}{{\cal A}/{\cal G}}       
\newcommand{\g}{\gamma}                   
\renewcommand{\t}{\theta}
\newcommand{\vt}{\vartheta}
\renewcommand{\o}{\omega}
\renewcommand{\S}{\Sigma}
\newcommand{\Sg}{\mbox{$\Sigma _{g}$}}
\newcommand{\Sd}{\mbox{$\Sigma _{0,1}$}}
\newcommand{\Sc}{\mbox{$\Sigma _{0,2}$}}
\newcommand{\Sp}{\mbox{$\Sigma _{0,3}$}}
\newcommand{\M}{{\cal M}_{F}(\Sg , G)}     
\newcommand{\del}{\partial}
\newcommand{\G}{\hat{G}}                    
\newcommand{\ex}[2]{e^{-e^{2}c(#1)A(#2)/2}} 
\newcommand{\ot}{\otimes}                   
\renewcommand{\theequation}{\thesection.\arabic{equation}}

\section{Introduction}
Topological gauge theories were the first of the topological field
theories to be put forward. The two broad types of topological field
theory were introduced originally by Schwarz \cite{schwarz}, to give a
field theoretic description of the Ray-Singer torsion \cite{rst}, and by
Witten \cite{wtft}, to give path integral representations of the Donaldson
polynomials \cite{d}.

Of the Schwarz type, only the Chern-Simons model of Witten has been
extensively analysed \cite{wjp}. The non-Abelian generalizations of
Schwarz's original actions, the so called $BF$ models \cite{bt1,h} have
been shown to have
partition functions which reduce to integrals over the moduli space
of flat connections with some power of the Ray-Singer torsion as the
measure \cite{bt1,bt2}. Apart from establishing that certain
correlators calculate intersection numbers of submanifolds there have been
virtually no concrete calculations performed with these theories
\cite{bt1,hs}.

The Witten or cohomological type theories have suffered a similar fate.
The one exception here being two dimensional topological gravity
\cite{wtpg} where a wide range of interesting results have been
obtained (see \cite{dijk} for a recent review).
On the
formal side, topological gauge theories, of Witten type, can be associated
with particular geometric structures on the space of connections $\A$
modulo the group of gauge transformations ${\cal G}$. $\AG$ has a natural
principle bundle structure (the universal
bundle of Atiyah and Singer \cite{as}) and also a natural Riemannian
structure \cite{nr}-\cite{gp}. Hitherto, in cohomological gauge theories,
$\AG$ has been considered from the principle bundle point of view
\cite{bs}-\cite{kanno} and as a Riemannian manifold \cite{bta,btb}.
For a general reference to, both the Witten and Schwarz, topological
theories see \cite{bbrt}.

There is one more geometric structure that may be placed on $\AG$, under
ideal circumstances, and this particular aspect of the space is a
meeting ground for the Schwarz and Witten type theories. Depending on the
underlying manifold $M$ it may be possible to induce a symplectic
structure on $\A$. This is indeed possible when $M$ is K\"{a}hler
(though it need not be). Examples include Riemann surfaces and complex
K\"{a}hler surfaces.

The topological field theories, that we will be concerned with, are
a topological gauge theory of flat connections over Riemann surfaces
and a topological gauge theory of instantons over four
dimensional manifolds. It turns out that, in order to define the
topological theory, one needs to `regularize' the model to avoid
problems with reducible connections. This regularization amounts to
considering instead Yang-Mills theory which, in the limit as
the gauge coupling $e^{2}$ goes to zero, reduces to the topological theory
\cite{w2d,w2dr}.
We find ourselves in the interesting situation of studying a `physical'
theory in order to extract topological information. Indeed most of the
lectures are devoted to an evaluation of the path integrals of
Yang-Mills theory on Riemann surfaces.

Now, in their own right, gauge theories in two dimensions have for a long
time served as useful laboratories for testing ideas and gaining insight
into the properties of field theories in general. While classically
Yang-Mills theory on topologically non-trivial surfaces is well
understood \cite{ab}, very little effort had gone into understanding
quantum gauge theories on arbitrary Riemann surfaces, the notable
exception being in the context of lattice gauge theory \cite{R} which is
based on previous work by Migdal \cite{M} (see also \cite{wheater}).
In the continuum quantum Yang-Mills theory on ${\bf R}^{2}$ was solved
in \cite{M,bralic} and on the cylinder in \cite{raj}.

Here we study Yang-Mills theory from the path integral point of view.
In particular we will get general and explicit expressions for the
partition function and the correlation functions of (contractible and
non-contractible) Wilson loops on closed surfaces of any genus as well as
for the kernels on surfaces with any number of handles and boundaries.
These expressions will yield corresponding results for the topological
theory in the limit. We will not be able to fix overall constants in our
formulae, these require a more detailed analysis and/or input from another
source. The method of calculation is based on published work
with Matthias Blau \cite{bt}. An analogous, but perhaps more
mathematically rigorous, derivation of some of these results may be found
in the work of Fine \cite{fine}. There are also unpublished lectures by P.
Degiovanni \cite{deg} where a mixture of canonical quantization and the
axiomatic approach to topological field theories is used to get to these
results. A derivation in the spirit of \cite{M,R}, was provided by
Witten \cite{w2d}.

Perhaps the correct way of of deciding on the `type' of topological field
theory one has in hand is with respect to which fixed point theorem applies
to it. Atiyah
and Jefrey \cite{aj} have shown that the cohomological field theories,
as they had been discussed, were naturally understood in terms of the
Marhai-Quillen construction \cite{mq}. An introductory account of this point of
view, explaining how the zeros of a map are singled out, is given in
\cite{b}. On the other hand, it had also been known that the path integral
formulation of index theorems \cite{ab2} devolved to calculations of
fixed points because of the theorem of Duistermaat and Heckman
\cite{dh}. It is this aspect of the two dimensional cohomological gauge
theory that is stressed in \cite{w2dr}. Unfortunately, there is no time
to go into this side of things, except in passing.

I have taken this opportunity to prove some of the technical facts that
were passed over in \cite{bt} and also to include some previously unpublished
calculations \cite{btym2}. Notation is by and large explained in appendix
\ref{note}.

\setcounter{equation}{0}
\section{Moduli Space Of Flat Connections And Topological Gauge Theory}

Our basic concern in these notes is with the space of flat connections
(gauge fields) on a Riemann surface. We define this space on a general
manifold $M$. Pick a connected, compact gauge group $G$.
A connection $A$ on a $G$ bundle over $M$, or a gauge field on $M$, is
said to be flat when its curvature tensor $F_{A}$ vanishes,
\be
F_{A} =dA + \frac{1}{2}[A,A] = 0 \, . \label{flat}
\ee
Flatness is preserved under gauge transformations $A \rightarrow A^{U}$ where
\be
A^{U} = U^{-1}AU + U^{-1}dU \, , \label{gt}
\ee
as $F_{A}$ transforms to $U^{-1}F_{A}U$. The moduli space of flat
connections ${\cal M}_{F}(M,G)$ is the space of gauge inequivalent
solutions to (\ref{flat}). This means
that solutions to (\ref{flat}) which are not related by a gauge
transformation are taken to be different points of ${\cal M}_{F}(M,G)$. On
the other hand,
if two solutions to (\ref{flat}) are related by a gauge transformation
they are taken to be the same point in ${\cal M}_{F}(M,G)$, that is
$A^{U} \equiv A$.

There is another description of the moduli space which is useful. This
is in terms of representations of the fundamental group $\pi_{1}(M)$ of
the manifold $M$,
\be
{\cal M}_{F}(M,G) = Hom(\pi_{1},G)/G \, , \label{hom}
\ee
that is of equivalence classes of homomorphisms
\be
\vf :\pi_{1}(M) \rightarrow G
\ee
up to conjugation. $\pi_{1}(M)$ is made up of loops on the manifold $M$
with two loops identified if they can be smoothly deformed into each
other. All contractible loops are identified. $\pi_{1}(M)$ is a group
under the composition of loops with the identity element the
contractible loops.

We can easily see half of (\ref{hom}). Given a flat
connection $A$ we can form a map $\vf_{\g}$ by setting $\vf_{\g}(A)$ to
be the holonomy (which is an element of the group $G$) around a loop $\g$
in $M$,
\be
\vf_{\g}(A) = P \exp{\int_{\g} A} \, , \label{vf}
\ee
In physicists notation this is a Wilson loop. Recall that $P$ stands for
path ordering
\be
P \exp{\left(\int_{0}^{1} f(t) dt\right)} = \sum_{n=0}^{\infty}
\int_{0}^{1}dt_{1 }\int_{0}^{t_{1}}dt_{2 }\dots \int_{0}^{t_{n-1}}dt_{n
} \, f(t_{n}) \dots f(t_{1}) \, . \label{path}
\ee
Under a gauge transformation (\ref{vf}) goes to
\be
\vf(A^{U}) = U(0)^{-1}\vf(A)U(1) \, ,
\ee
so, as $U(0)=U(1)$, gauge equivalent $A$'s give conjugation equivalent
$\vf(A)$'s.

We still need to show that the maps only depend on the
homotopy class of the loop $\g$. This is where flatness comes in; we
have not used it yet. Add to $\g$ a small homotopically trivial loop $\d
\g= \del \Gamma$ (it is the boundary of some disc $\Gamma$) then
\bea
& & \vf_{\g + \d \g}(A) -\vf_{\g}(A) \nonumber \\
& & = \int_{0}^{1}dt P\exp{\left( \int_{0}^{t}A_{\m}\frac{d
\g^{\m}(s)}{ds} ds \right)}\, A_{\m}(t) \frac{d \d \g ^{\m}(t)
}{dt} \, P\exp{\left( \int_{t}^{1}A_{\m}\frac{d
\g^{\m}(s)}{ds} ds \right)} \nonumber \\
& & =  \int_{0}^{1}dt P\exp{\left( \int_{0}^{t}A_{\m}\frac{d
\g^{\m}(s)}{ds} ds \right)}\, F_{\n \m}(t) \frac{d\g^{\n}(t)}{dt} \d
\g^{\m}(t) \, P\exp{\left( \int_{t}^{1}A_{\m}\frac{d
\g^{\m}(s)}{ds} ds \right)} \nonumber \\
& & =0 \, .
\eea
The first equality follows from the variation of the definition of path
ordering (\ref{path}), while the second arises on integrating by parts
in $t$. We have just shown that only the homotopy class of the loop $\g$
is involved in the map $\vf_{\g}$.

This establishes that each point in ${\cal M}_{F}(M,G)$ gives an element in
$Hom(\pi_{1}(M) ,$ $G)/G$. The proof of the converse, that each element in
$Hom(\pi_{1}(M),G)/G$ naturally defines a flat connection, makes use of
the notions of covering manifolds and associated bundles.

\noindent \underline{Dimension Of $\M$}

We now concentrate on compact Riemann surfaces of genus $g$, $M=\Sg$, and
compact gauge group $G$. In this case it is known that $\M$ is smooth
except at singular points which arise at reducible connections. The
reducible connections will be defined shortly; a great deal of the
formalism developed is there to get around problems generated by these
connections.

Now to a Riemann surface $\Sg$ there is a standard
presentation of $\pi_{1}$ in terms of the $2g$ generators $a_{i}$,
$b_{i}$, $i=1, \dots, g$. One basis for these `homology' cycles is
displayed in figure $1$. However, they are not independent generators.
To see this it is easiest to form the cut Riemann surface. One picks a
point $P$ on the surface and then cuts from that point along a
fundamental cycle
back to the point. This is repeated for the basis of cycles, the final
result being a cut Riemann surface. In figure $2$ this process is shown
for the torus $T^{2}$. Figure $3$, shows the homology basis for the genus
$2$ surface, and how the basis is pulled to the point $P$ and cut is
shown in figure $4$. Now
the path that is defined by the edge of the cut Riemann surface is generated
by
\be
a_{1}b_{1} a_{1}^{-1}b_{1}^{-1}\dots a_{g}b_{g} a_{g}^{-1}b_{g}^{-1} \,
,
\ee
but this path is contractible to a point in the interior of the cut Riemann
surface, so it is the trivial element in $\pi_{1}$. We have the relation
\be
a_{1}b_{1} a_{1}^{-1}b_{1}^{-1}\dots a_{g}b_{g} a_{g}^{-1}b_{g}^{-1}=1
\, . \label{rel}
\ee
It turns out that this is the only relation satisfied by the generators
on the Riemann surface.

The dimension of the moduli space for $g>1$ and simple $G$ may be
calculated from the information that we have at hand. The $Hom$ part of
(\ref{hom}) asks for the possible assignment of group elements to
generators. There are $2g\, dim\,G$ ways of doing this, but we must subtract
off the one relation (\ref{rel}), that is minus $dim\, G$ and also the
identification
of conjugacy clases implies that we ought to subtract another $dim\,G$. We
have, therefore,
\be
dim \, \M  = (2g-2)\,dim\,G \, .
\ee
When the manifold is the two sphere, $g=0$, all loops are contractible
so $\pi_{1}(S^{2}) =id$ and $\M$ is one point, the trivial representation.
This means that up to gauge equivalence the only flat connection is the
trivial connection $A=0$. For the torus, $g=1$, the situation changes
somewhat. In this case the relation (\ref{rel}) is $ab=ba$ so that $a$
and $b$ must commute. The homomorphism must therefore ensure that when
mapped into $G$ their images commute. Generically $a$ and $b$ can be
represented, in this case, by elements lying in the (same) Maximal torus
$T$ of $G$. The dimension is
\be
dim \, {\cal M}_{F}(\S_{1},G) = 2\, dim \,T \, .
\ee

Life is simplified when $G=U(1)$.  As everything in sight must commute,
the relation (\ref{rel}) is automatically satisfied and conjugation acts
trivially. We have $dim \, {\cal M}_{F}(\Sg, U(1) ) = 2g$.

\noindent \underline{Topological Gauge Theory}

We would like to be able to get more information than just the
dimension. Different types of topological field
theories indeed give different sorts of information about these moduli
spaces. Let us define what we mean by a topological field theory.

For the purposes of these lectures a topological field theory is a field
theory defined over some manifold $M$ whose partition function is invariant
under smooth deformations of any metric one puts on $M$. In such a
theory it is possible to find correlation functions which enjoy the same
property. A topological gauge theory is a topological field theory which
is also a gauge theory. The correlation functions of interest in this
case need to be not only metric independent but also gauge invariant.

In the course of the lectures we will come across two types of
topological gauge theory. The first, known as a $BF$ model, has a
partition function that equals the volume of $\M$. Due to the
singularities of the moduli space, we will need to generalise the
discussion somewhat and consider Yang-Mills theory. The partition
function for the Yang-mills theory will be determined and in the
topological limit we will be able to get a handle on $vol \, \M$. The
second type of topological field theory that we come across is known as
a cohomological gauge theory. Considerations from this theory show us
that the volume we calculate is the symplectic volume of $\M$.
Correlation functions in the cohomological theory may be interpreted in
terms of intersection theory on the moduli space. We consider only the
dual point of view, that is as integration of differential forms over
$\M$.

\setcounter{equation}{0}
\section{$BF$ Theory on a Riemann Surface}
We are interested in the moduli space of flat connections ${\cal
M}_{F}(\Sg , G)$ on a Riemann surface of genus $g$ and compact structure
group $G$. In the previous section we saw that this is the space of
gauge inequivalent solutions to the flatness condition
\be
F_{A} = 0 \, .
\ee

A field theory that restricts one to this space is given by the path integral
\cite{bt1,h}
\be
Z(\Sg ) = \int D \f DA \exp{\left( \frac{1}{4\pi^{2}}\int_{\Sg }Tri \f F_{A}
\right)  } \, , \label{bf}
\ee
where $\f$ is an adjoint valued field. Traditionally the field $\f$
here is denoted by $B$ and a glance at the partition function will
explain the reason for the name $BF$ theory. We have broken with
tradition in order to make a smooth transition to the cohomological model.
Formally, at least, on integrating
out $\f$ the path integral gives the volume of the moduli space of flat
connections
\bea
Z(\Sg) & =& \int DA \delta(F_{A}) \nonumber \\
           & = & vol\,{\cal M}_{F}(\Sg , G) \, .
\eea
As it stands this formula is far too implicit and we will have need to
modify it in making sense of the last equality. But first let us
establish some formal aspects of the theory.

$(i)$ Gauge Invariance:

The action in (\ref{bf}) is invariant under the gauge
transformation (\ref{gt}) combined with
\be
\f ^{U} = U^{-1} \f U \, .
\ee
The infinitesimal form of the transformations are
\be
\d _{\Lambda} A = d_{A} \Lambda , \; \; \; \; \d _{\Lambda} \f = [\f,
\Lambda] \,. \label{gt1}
\ee
In order to correctly specify the path integral, we will need to gauge fix.
The reason for this is that gauge invariant operators are constant on
the orbit of the group of gauge transformations. One gets an infinity as
one integrates over each orbit. It is this infinity that needs to be
factored out. Rather than integrating over $\A$ one wants to integrate
over $\AG$. We use the Fadeev-Popov method to pick the gauge and fix on
\be
G(A)=0 \,
\ee
where $G(A)$ could be, for example, $d_{A_{0}}*(A-A_{0}) $ (where $*$ is
the Hodge duality operator with respect to some metric $g_{\mu \nu}$ on $\Sg
$ and $A_{0}$ is some prefered connection) or, as
we will be mostly working on the disc, $ G(A)= A_{r}$. The partition
function is now
\be
Z_{\Sg}=\int  DA D\f Dc D\bar{c} Db
\exp{\left(\frac{1}{4\pi^{2}}\int_{\Sg}Tri\f F_{A}  + \int_{\Sg}Tr ( ib G(A) +
\bar{c} \frac{\d G}{\d A}d_{A}c ) \right) } \, . \label{bfa}
\ee
The extra contributions to the action may be written as a BRST variation
\be
Q \int_{\Sg}Tr \bar{c} G(A) \, ,
\ee
with $Q$ the BRST operator,
\bea
Q A = d_{A}c  \, , \; \; \; \; \;   & & Q c = -\frac{1}{2}[c,c] \, ,
\nonumber \\
Q \bar{c} = ib \, , \; \; \; \; \;  & & Q b = 0 \, .
\eea
As usual one has traded overall gauge invariance for BRST invariance.

This is not quite as much of the group volume that can be factored out.
Elements $h$ in $G$ (that is constant maps $h \in {\cal G}$) that form
the centre of $G$, $Z(G)$ do not act on $A$ or $\f$. It is also possible
factor out the number of elements $\# Z(G)$ so that one should consider
\be
\frac{1}{\# Z(G)}Z_{\Sg} \, ,
\ee
but this factor will generally be omitted.

$(ii)$ Metric Independence:

At the level of (\ref{bf}), this is manifest, for the
metric makes no appearance at all there. However, upon gauge fixing, we
have introduced an explicit metric dependence in the action of
(\ref{bfa}). All of the explicit metric dependence rests in $G(A)$, so
that, on varying (\ref{bfa}) with respect to the metric, we find
\be
\frac{\d Z_{\Sg}}{ \d g_{\mu\nu} } =  \int_{\Phi} e^{L(\Phi)}\, Q \,Tr
\bar{c}  \frac{\d G(A)}{ \d g_{\mu\nu} }  \, , \label{metv}
\ee
where $\Phi$ is generic for all the fields and
\be
L(\Phi) = \frac{1}{4\pi^{2}}\int_{\Sg}Tri\f F_{A}  + \int_{\Sg}Tr ( ib G(A) +
\bar{c} \frac{\d G}{\d A}d_{A}c ) \, .
\ee
By the BRST invariance of the theory the right hand side of
(\ref{metv}) vanishes, whence the metric independence of the partition
function is established (an account of how one derives such a Ward
identity is given in section \ref{coho}). This has all been rather formal. A
more careful analysis, working with a regularized form of the theory,
shows that indeed the theory remains metric independent, substantiating the
analysis we have made.

$(iii)$ Relationship to the Ray-Singer Torsion:

Let us suppose that the only flat connection is isolated and call it
$A_{0}$. Split the general connection $A$ into $A= A_{0} + A_{q}$ and
take as the gauge condition $d_{A_{0}}*A_{q}=0$. The path integral is
\bea
& & \int DA_{q} \d \big(F_{A_{0}+A_{q}}\big)\, \d \big(d_{A_{0}}*A_{q}\big) \,
\det{\big( d_{A_{0}}*d_{A_{0}+A_{q}} \big)} \nonumber \\
& & = \int DA_{q} \d \big(d_{A_{0}}A_{q}\big)\, \d \big(d_{A_{0}}*A_{q}\big) \,
\det{\big( d_{A_{0}}*d_{A_{0}+A_{q}} \big)}  \nonumber \\
& & = \det{\big(d_{A_{0}},d_{A_{0}}*\big)}^{-1}
\det{\big( d_{A_{0}}*d_{A_{0}} \big)}  \, , \label{rst}
\eea
where the last equality arises on noting that the two delta functions imply
$A_{q}=0$, and the inverse determinant comes from extracting the operators
out of the delta function with the rule
\be
\int_{-\infty}^{+\infty} dx_{1}\dots \int_{-\infty}^{+\infty}dx_{n}
\d\big(T(x) \big) = \det{\big(T(0)\big)}^{-1} \, .
\ee
Now the product of determinants is the Ray-Singer torsion. The torsion
is unity on an even dimensional manifold, a fact that is easy to prove
using field theoretic techniques \cite{bt1}, and we will do so for this
case when we consider the trivialising map. Path integral representations of
the
torsion were introduced by Schwarz \cite{schwarz}.

When the flat connections are not isolated the connection $A_{0}$ will
depend on `moduli' $\l$. The path integral must now include an
integration over the moduli parameters, but for any $\l$ the product of
determinants is still one, so that we are again, formally, left with
\be
\int d\l = vol \, \M \, .
\ee

\noindent \underline{Reducible Connections}

So far we have concentrated on the flatness equation (\ref{flat}) which
is one of the equations of motion that is obtained from the action of
(\ref{bf}). The other equation of motion, obtained on varying the action
with respect to the gauge field is,
\be
d_{A}\f = 0 \, . \label{red}
\ee
Connections $A$ for which there are non-zero solutions $\f$ to
(\ref{red}) are called reducible. Thinking of the $\f$ as gauge
parameters, then (\ref{red}) is the statement that there are some gauge
transformations that act trivially on the connection $A$. This means
that ${\cal M}_{F}(\Sg , G)$ is not, in general, a manifold as the
quotienting out by the gauge group is not the same at each connection.
Generically the connections are irreducible, and there will be isolated
reducible connections. ${\cal M}(\Sg , G)$ is then at best an orbifold. Turning
this into a bona fide manifold is the process of `compactification'.

(\ref{red}) clearly holds when the two conditions
\be
d\f = 0 \, , \; \; \; \; \; \; \; \; [A,\f] =0 \, ,
\ee
are fullfilled. As an example consider the $su(2)$ valued gauge field
\be
A = \left( \begin{array}{c}  a   \\ 0   \end{array}
\begin{array}{c} 0 \\ -a \end{array} \right) \, , \label{redc}
\ee
with the possible form of $\f$ being
\be
 \left( \begin{array}{c}  b   \\ 0   \end{array}
\begin{array}{c} 0 \\ -b \end{array} \right)  \, , \label{redb}
\ee
with $b$ a constant. The connection (\ref{redc}) and the scalar field
(\ref{redb}) live in a $u(1)$ subalgebra of $su(2)$. The $SU(2)$ gauge
field will be flat when $a$ is flat as a $U(1)$ gauge field.

Reducible connections are a source of great difficulty in making sense of
topological field theories in general. The problem is that at a
reducible connection path integrals of the type (\ref{bf}) diverge. The
reason for this is that there are integrals to be performed over all the
$\f$ modes, but those modes which satisfy (\ref{red}) do not appear
in the action and hence do not dampen the integrals. For the reducible
connection (\ref{redc}) there will be the undamped integrals
\be
\int_{-\infty}^{+\infty} db = \infty \, .
\ee

\noindent \underline{Yang-Mills Connections}

In order to overcome the problems associated with reducible connections,
Witten has suggested a way of `thickening' things out \cite{w2d}. The
idea is to spread the delta function (\ref{bf}) into a Gaussian in a
gauge invariant way. The partition function is taken to be
\be
Z_{\Sg}(e^{2}A(\Sg)) =\int DA D\f
\exp{\left(\frac{1}{4\pi^{2}}\int_{\Sg}Tr i\f F_{A} +
\frac{e^{2}}{8\pi^{2}}\int_{\Sg} Tr \f*\f \right)} \, . \label{y1a}
\ee
The ghosts and multiplier fields are implicit in this formula.

The
dependence on the coupling $e^{2}$ and the area of the surface $A(\Sg) =
\int *1$ in the combination $e^{2}A(\Sg)$ may be derived as follows.
Scale the metric by $g_{\m ,\n} \rightarrow \l g_{\m , \n}$ then the
term
\be
\int_{\Sg} Tr \f * \f = \int_{\Sg} \sqrt{detg} Tr \f \f \, ,
\ee
scales to
\be
\l \int_{\Sg} Tr \f * \f \, .
\ee
This factor may be eliminated if we in turn
send $e^{2}$ to $\l^{-1}e^{2}$. Shortly we will see the existence of a
map that guarantees that the metric only enters as a measure (there are
no derivatives of it). The invariant combination is then $e^{2}A(\Sg)$.

For an arbitrary metric, with no loss of
generality, the part of the action $ e^{2} \int_{\Sg} Tr \f * \f $ may
be replaced with $e^{2}A(\Sg)\int_{\Sg} Tr \f * \f $ where the metric
here has area fixed to unity, $ \int_{\Sg}*1 =1$. Because of this, we
adopt the following convention: in the the formulae obtained for the
evaluation of the path integral the combination $e^{2}A(\Sg)$ will be
denoted by $\ep$, but also in the action we set $\ep =e^{2}$.

This is a rewriting of the Yang-Mills path integral which makes the
relation to the topological theory of flat connections transparent. To
see that this is the same as the Yang-Mills partition function,
perform the Gaussian integration over the field $\f^{a}$ with
\be
\int_{-\infty}^{+\infty}\frac{dx}{\sqrt{2\pi}} \exp{(- e^{2}x^{2}/2
+ixy)} = \frac{1}{\sqrt{e^{2}}}\exp{(- y^{2}/2e^{2})} \, , \label{gaus}
\ee
to obtain
\be
Z_{\Sg}(\ep) =\int DA
\exp{\left(+ \frac{1}{8\pi^{2} \ep }\int_{\Sg}Tr F_{A}*F_{A} \right)} \, .
\ee
The original theory is obtained in the $e^{2} \rightarrow 0$ limit (or
in the limit $A(\Sg) \rightarrow 0$).

The minima of the new action in (\ref{y1a}) are
\be
F_{A} = -i e^{2} * \f \, , \; \; \; \; d_{A}\f = 0 , \label{yma}
\ee
on combining the two one obtains the Yang-Mills equations
\be
d_{A}*F_{A} = 0  \label{ym}
\ee
A gauge field satisfying (\ref{ym}) is said to be a Yang-Mills
connection. The solutions to (\ref{yma}) fall into two distinct classes.
The first is that $\f=0$ in which case the connections are flat. At this
point we see that there is
a partial resolution of the problem facing us, for in this sector we now
have no $\f$ zero-modes to worry about. This does not mean that the flat
connections can not be reducible. There may well be non-zero Lie algebra
valued functions $\vf$ for these connections such that $d_{A}\vf = 0$.
The equations tell us that these $\vf$ are not proportional to $F_{A}$.
The second class has $\f \neq 0$ for which the connections are
not flat but are certainly reducible. As $e^{2} \rightarrow 0$ these
classes merge to give back the complicated situation of flat connections
of which most, but not all, are irreducible.

To see that the action ``regularizes'' the contribution to the path
integral of elements of the isotropy subgroup of the group of gauge
tansformations consider the pair (\ref{redc},\ref{redb}) and take $a$ to
be flat (choose the bundles so that this is possible). Inserting
these into the path integral (\ref{y1a}) we see that up to the volume of
the flat $u(1)$ connections we are left with an integral
\be
\int_{-\infty}^{+\infty} \frac{db}{\sqrt{2\pi }} \exp{\left(-\frac{1}{4
\pi^{2}}\ep
b^{2}  \right) } =
\sqrt{2\pi^{2}/\ep} \, ,
\ee
thus regularizing the infinity obtained on using the original path
integral.

\noindent \underline{A Trivialising Map}

There is no dynamics in pure Yang-Mills theory in two dimensions, even at
the quantum level, for there are no physical degrees of freedom
associated with the gauge field\footnote{In $d$ dimensions the number of
physical polarizations of a gauge boson is $d-2$.}. Indeed we will present
a map which eliminates all (local) reference to the differential operators
that are implicit in (\ref{y1a}). Explicitly the partition
function is now
\bea
Z_{\Sg}(\ep) & =& \int  DA D\f Dc D\bar{c} Db
\exp{\left(\frac{1}{4\pi^{2}}\int_{\S}Tri\f F_{A} +
\frac{\ep}{8\pi^{2}}\int_{\S} Tr \f*\f  \right. } \nonumber \\
& & \left. \; \; \; \; \; \;  + \int_{\S}Tr ( ib G(A) +
\bar{c} \frac{\d G}{\d A}d_{A}c ) \right) \, . \label{eq:y2a}
\eea
The map that we have in mind is $A \rightarrow (\xi,\eta)$ defined by
\cite{bt1,bt2}
\bea
\xi(A) &=& F_{A} \; \; , \nonumber \\
\eta(A) &=& G(A) \; \;  , \label{eq:nicmap}
\eea
which has as its (inverse) Jacobian
\be
J^{-1} = det \frac{\delta(\xi,\eta)}{\delta A} = det (d_{A},\frac{\delta
G(A)}{\delta A}) \, .
\ee
Taking $G(A)= d_{A_{0}}*(A-A_{0})$, we need to determine
\be
det(d_{A}, d_{A_{0}}*) = det T \, .
\ee
Here $T$ may be thought of as the map
\bea
& & T : \Omega^{1}(\Sg, Lie G) \rightarrow \Omega^{0}(\Sg, Lie G) \oplus
\Omega^{0}(\Sg, Lie G) \nonumber \\
& & \; \; \; \; T(\alpha) = ( *d_{A}\alpha , *d_{A_{0}}* \alpha) \, .
\eea
We can give a path integral representation of this as
\be
det T = \int D\sigma D\bar{\sigma} D\alpha \exp{\left( i Tr \int_{\Sg}
\sigma d_{A}\alpha +
\bar{\sigma}d_{A_{ 0}}* \alpha \right)} \, ,
\ee
where $\sigma$ and $\bar{\sigma}$ are Lie algebra valued Grassmann odd
functions, and
here $\alpha$ is a Lie algebra valued Grassmann odd one form. In order to
get a handle on the determinant we define
\bea
det_{\ep} T& =& \int D\sigma D\bar{\sigma} D\alpha \exp{\left(  Tr
\int_{ \Sg}  i\sigma d_{A} \alpha + i\bar{\sigma}d_{A_{ 0}}* \alpha  +
\frac{\ep}{2} \alpha  \alpha \right)}  \nonumber \\
& = & \int D(\sigma /\ep) D\bar{\sigma} \exp{\left(- Tr \int_{\Sg} \,
\frac{1}{2\ep}(d_{A}\sigma  + *d_{A_{0}}\bar{\sigma}) *
(d_{A}\sigma   + *d_{A_{0}}\bar{\sigma})
\right)} \nonumber \\
& = & \int D\sigma D\bar{\sigma} \exp{\left(- Tr\int_{\Sg} \,
\frac{1}{2\ep}(\ep d_{A}\sigma  + *d_{A_{0}}\bar{\sigma})
(\ep  d_{A}\sigma  + *d_{A_{0}}\bar{\sigma}) \right)} \, , \label{det}
\eea
with
\be
det T \equiv det_{0}T \, .
\ee
Taking the limit $\ep \rightarrow 0$ in (\ref{det}) is
straightforward
\be
det T  = \int D\sigma D\bar{\sigma} \exp{\left(-Tr \int_{\Sg}\,
 \bar{\sigma} d_{A_{0}}*d_{A}\sigma \right)} \, ,
\ee
but this is precisely the ghost determinant. The Jacobian is then seen
to exactly cancel the determinant that arises from integrating out the
Fadeev-Popov ghosts $(c,\bar{c})$. Other gauge choices may be dealt with
in the same manner. In passing we note that this definition of
determinants is equivalent to the definition in \cite{schwarz}. We
have also established that the ratio of determinants (\ref{rst}) that make
up the Ray-Singer torsion is unity.

In this way the path integral seems to be
\be
Z_{\Sg}(\ep)=\int   D\f D\xi D\e Db \exp{\left( \frac{1}{4 \pi^{2}}
\int_{\Sg}Tr i\f \xi  + \frac{\ep}{8\pi^{2}}\int_{\Sg} Tr \f*\f + \int_{\Sg}
Tri b \e  \right)} \, , \label{eq:y3a}
\ee
which is formally one. This is not quite correct as we have not encoded the
global properties of the map into the path integral yet. The dependence
on $\ep$ factors, as in the second line of (\ref{det}), if the number of
modes of one of the components of $\a$ and the number of modes of
$\sigma$ agree. These agree up to global mismatches. An example is
furnished by the reducible connections. Here there will be $\sigma$ and
$\bar{\sigma}$ zero modes making (\ref{det}) ill defined. The
determinant will also be singular on the set where $d_{A} \alpha =0$ and
$d_{A_{0}}*\alpha = 0$ so that $det_{0}T$ may vanish, that is, these modes
are not
weighted in the path integral. If they lie on a compact space then they
contribute a finite volume factor (as we expect about flat connections),
otherwise one needs to take expectation values which explicitly damp the
integrals.

For a surface
with boundary, the boundary value of the connection needs to be specified
in (\ref{eq:y2a}). Consequently this data must be expressed in terms of
$\xi$ and $\eta$ in (\ref{eq:y3a}). For a surface without boundary there
are also topological constraints on $F_{A}$ and hence on $\xi$. In the
following sections we  will see how to incorporate these global aspects
of gauge theories on Riemann surfaces.

In any case, questions of global modes aside, we have established that
the metric enters only as a measure and consequently the coupling
constant indeed always appears in the combination
$e^{2}A(\Sg)$.

\noindent \underline{Observables}

The natural topological observables in the $BF$ theory defined by
(\ref{bf}) are Wilson loops around non-contractable cycles $\gamma$ in
some representation $\l$ of $G$
\be
W[\gamma,\l] = Tr_{\l} P \exp{\left( \oint_{\gamma} A \right)} \, ,
\ee
and Wilson points
\be
W[x,\l, q] = Tr_{\l} \exp{\left(q \f (x) \right)} \, .
\ee
These clearly do not depend on any metric and are gauge invariant. We
can show that the expectation values of products of these observables depend
only on the homotopy class of the cycles and not at all on the points (as long
as they do not touch each other). Note that
\bea
d W[x,\l, q]& =&  Tr_{\l} d\exp{\left( q \f(x) \right)} \nonumber \\
&=& Tr_{\l}\left( \right. \exp{\left( q \f(x) \right) } qd \f \left.
\right) \nonumber \\
&=& Tr_{\l}\left( \right. \exp{\left( q \f(x) \right) } qd_{A} \f \left.
\right) \, .
\eea
The vacuum expectation value of $W[x,\l]$ is written as
\be
<W[x,\l]> = \int_{\Phi} W[x,\l] \exp{\left(
\frac{1}{4\pi^{2}}\int_{\S}Tri \f F_{A} \right)} \, ,  \label{wpbf}
\ee
so that on differentiating with respect to the point $x$ we find
\bea
d <W[x,\l]>& =& \int_{\Phi} d W[x,\l] \exp{\left(
\frac{1}{4\pi^{2}}\int_{\S}Tri \f F_{A} \right)} \nonumber \\
&=& \int_{\Phi} Tr_{\l}\left( \right. \exp{\left( q \f(x) \right) } qd_{A}
\f  \left. \right) \exp{\left(
\frac{1}{4\pi^{2}}\int_{\S}Tri \f F_{A} \right)} \nonumber \\
 &=& \int_{\Phi} Tr_{\l}\left( \right. \exp{\left( q \f(x) \right) } q 4i
\pi^{2} \frac{\d}{\d A}   \left. \right) \exp{\left(
\frac{1}{4\pi^{2}}\int_{\S}Tri \f F_{A} \right)} \nonumber \\
&=& 0 \, .
\eea
In the last line we used the fact that the path integral over a total
divergence in function space is zero.

A similar exercise shows the homotopy invariance of $<W[\gamma,\l]>$. If
we vary $\gamma$ by adding a small loop $\d \gamma = \partial \Gamma$
then we have
\bea
& & \d <W[\g,\l]> \nonumber \\
& & = \int_{\Phi} Tr_{\l}P\left( \right. \exp{\left( \oint_{\g}
A\right) } F_{A} d \Gamma \exp{\left( \oint_{\g}
A\right) } \left. \right) \exp{\left(
\frac{1}{4\pi^{2}}\int_{\Sg}Tri \f F_{A} \right)} \nonumber \\
& &= 0 \, .
\eea
The last line follows on integrating over $\f$.

A general expectation value will have the form
\bea
& & < \prod_{i=1}^{m} \prod_{j=1}^{n} W[\gamma_{i},\l_{i}] W[x_{j}, \m_{j}]
>  \nonumber \\
&=&  \int_{\Phi} \prod_{i=1}^{m} \prod_{j=1}^{n} W[\gamma_{i},\l_{i}] W[x_{j},
\m_{j}]  \exp{\left(
\frac{1}{4\pi^{2}}\int_{\Sg}Tri\f F_{A} \right) } \, ,
\eea
and will not depend on the deformations $\d \gamma_{i}$ or the points
$x_{k}$ providing one may organize for these never to intersect.

The Wilson points have an interesting consequence, in that they move us
away from flat connections. We represent the trace of the Wilson point
in terms of Grassmann anihilation $\bar{\e}$ and creation $\e$
operators,
\be
Tr_{\l}\exp{\left(q \f  \right)} = <0\mid \bar{\e}^{i} \exp{\left(q \f^{a}
\e^{ m} \l_{mn}^{a}
\bar{\e}^{n} \right)} \e_{i}\mid 0>  \, , \label{wp}
\ee
with
\be
\{ \e_{i}, \bar{\e}_{j} \} = \d_{ij} \, , \; \; \{ \e_{i}, \e_{j} \} = 0
\, , \; \; \{ \bar{\e}_{i}, \bar{\e}_{j} \} = 0 \, ,
\ee
and
\be
<0 \mid 0> = 1 \, , \; \; \bar{\e}_{i} \mid 0> = 0 \, , \; \; <0 \mid
\e_{i} =0 \, .
\ee
With this representation we see that (\ref{wpbf}) takes its values at
\be
F_{A}^{a}(z) = q \e^{m} \l_{mn}^{a} \bar{\e}^{n} \d^{2}(z-x) \, ,
\label{aflat}
\ee
so that away from the point $x$ the connection is flat but at the point
it has a curvature with delta function support. If there are more Wilson
point operators inserted into the path integral then the right hand side
of (\ref{aflat}) becomes a sum
\be
F_{A}^{a}(z) = \sum_{i=1}^{n} q_{i} (\e_{i})^{m} (\l_{i})_{mn}^{a}
(\bar{\e}_{i})^{n} \d^{2}(z-x_{i}) \, .
\ee

If one considers the expectation value of operators of the form
$\exp{\left(q  \f(x) \right)}$ then gauge invariance is lost at the
point $x$. Calculations of this type correspond to considering `pointed'
gauge transformations, that is those that do not act at the prefered points
and one talks of `marked' Riemann surfaces.

\setcounter{equation}{0}
\section{Cohomological Field Theory} \label{coho}
In this section we will give an explanation as to the ``type'' of volume
being calculated for $\M$. In order to do this we
introduce the basic ideas behind topological gauge theories of
cohomological, or Witten, type. These were originally proposed by Witten
to give a field theoretic description of the Donaldson polynomials.
These metric invariants are defined as cohomology classes on the space of
anti-self-dual instantons over a given four manifold. The appropriate
framework for discussing these ideas from the path integral point of
view is in terms of the universal bundle introduced by Atiyah and Singer.

The set up is the following \cite{as}. One takes $P$ to be a principal
$G$ bundle over a manifold $M$, $\A$ the affine space of connections on $P$ and
${\cal G}$ the group of gauge transformations. There is a natural action
of ${\cal G}$ on $P \times \A$ with no fixed points so that $P \times
\A$ is a principle bundle over $(P \times \A)/{\cal G}= {\cal Q}$. There
is also a natural action of $G$ on ${\cal Q}$ so that away from
reducible connections (or for ${\cal G}$ the pointed gauge group) ${\cal
Q}$ is itself a principle $G$ bundle over $M \times \AG$. There is a
bigrading of differential forms on $M \times \AG$, a (p,q) form
is a $p$-form on $M$ and a $q$-form on $\AG$.

In topological gauge theory a set of fields appear that naturally
correspond to geometric objects in the universal bundle. These are the
gauge field $A$ (a $(1,0)$-form), a Grassman odd, Lie-algebra valued, one
form $\p$ (thought of as a $(1,1)$-form) and a Lie
algebra valued, Grassmann even, zero form $\f$ (a $(0,2)$-form). They
are related by the
BRST supersymetric transformation rules,
\be
\d A = \p \, , \; \; \; \d \p =  d_{A} \f \, , \; \; \; \d \f =0 \, .
\label{ttr}
\ee
If $\p$ is given a Grassmann charge of $1$ then $\f$ has charge $2$ and
these are their form degrees on $\AG$.
The geometrical interpretation is the following. $\d$ is viewed as the
exterior derivative in the `vertical' direction so that $\p$ is, by the
first equation, a curvature two form, a one form in the base manifold
(horizontal) direction and a one form vertically (this is why it
is Grassmann odd). $\f$ is a curvature two form in the vertical
direction. With this interpretation the last two equations in
(\ref{ttr}) are Bianchi identities.

\noindent \underline{Conventional Formulation}

We wish to model not the instanton moduli space but ${\cal M}_{F}(\Sg ,
G)$. In order to do this we introduce the fields $B$, $\chi$, $\bar{\f}$
and $\eta$ which are all Lie-algebra valued zero forms. They are,
however, Grassmann even, odd, even and odd respectively. Their BRST
transformation rules are
\bea
\d \chi = B \, , \; \; \; & & \d B = [\chi, \f] \, , \nonumber \\
\d \bar{\f} = \eta \, , \; \; \; & & \d \eta = [\bar{\f}, \f] \, .
\label{ttr2}
\eea
With these rules the BRST transformation on any field, $\Phi$, satisfies
\be
\d ^{2} \Phi = {\cal L}_{\f} \Phi \, ,
\ee
with ${\cal L}_{\f}$ being a gauge transformation with gauge parameter
$\f$.

The action is chosen to be \cite{bbt,ms}
\bea
L &= & \d \; i\int_{\Sg} Tr \left( \chi F_{A} + \bar{\f} d_{A}*\p \right)
\nonumber \\
& & \,  = i \int_{\Sg} Tr \left( BF_{A} - \chi d_{A} \p + \eta
d_{A} * \p  + \bar{\f} d_{A}*d_{A} \f + \bar{\f} \{ \p, * \p \} \right)
\, . \label{dbf}
\eea
As the integrand on the right hand side is gauge invariant, we see that
the application of $\d$ once more vanishes, that is the action is BRST
invariant. This action seems appropriate for our needs as the integral over
$B$ yields a delta function constraint onto the flat connections. That
it defines a topological theory is not quite apparent, for the Hodge
duality operator appears explicitly. Let $\tilde{Z}$ be the path
integral with this action. Then the metric variation $\d _{g}$ of
$\tilde{Z}$ is
\bea
\d _{g} \tilde{Z} &=&  \int_{\Phi} e^{L} \d _{g}L \nonumber \\
& =&  \int_{\Phi} e^{L} \d  \d _{g} \int_{\Sg} Tr
\left(  \chi F_{A} + \bar{\f} d_{A}*\p \right) \nonumber \\
& = &  \int_{\Phi} \d V_{g} \, ,  \label{metv1}
\eea
the last line being a defining equation for $V_{g}$. The order of $\d$
and $\d _{g}$ is not important as they commute (basically because the
transformation rules (\ref{ttr}) and (\ref{ttr2}) do not involve the metric).

The last line may be shown to vanish in some generality. Consider the
vacuum expectation value of any operator ${\cal O}$
\be
\int_{\Phi} e^{L(\Phi)} {\cal O}(\Phi) \, .
\ee
One may change integration variables $\Phi \rightarrow \Phi + \d \Phi$
and note that the action satisfies $L(\Phi + \d \Phi) = L(\Phi)$ while
also formally the path integral measure has the same property
$\int_{\Phi + \d \Phi} = \int_{\Phi}$. In terms of the new variables the
expectation value is
\be
\int_{\Phi} e^{L(\Phi)} {\cal O}(\Phi + \d \Phi) = \int_{\Phi}
e^{L(\Phi)} \left( {\cal O}(\Phi) + \d {\cal O}(\Phi) \right) \, ,
\ee
from which we conclude that
\be
\int_{\Phi} e^{L(\Phi)} \d {\cal O}(\Phi) = 0 \, .  \label{dexact}
\ee
This is exactly what is required to set the last line of (\ref{metv1}) to
zero. Indeed, replacing everywhere in this derivation $\d$ with $Q$
gives the Ward identity needed to establish that (\ref{metv}) vanishes,
as well.

The theory defined by $L$ seems to be just what we want, a topological
field theory that lands on $\M$. However, the
partition function $\tilde{Z}$ suffers greatly at the hands of the
reducible connections. For, at a reducible connection, there are zero
modes for the $B$, $\chi$, $\bar{\f}$, $\eta$ and $\f$ fields!

\noindent \underline{A Formulation In Terms Of The Symplectic Geometry
Of $\AG$}

Witten has proposed a method for avoiding the problems associated with
the reducible connections \cite{w2dr}. In this approach there is no need to
introduce the fields $B$, $\chi$, $\bar{\f}$ or $\eta$ at all. Rather,
one begins with the supersymmetric action
\be
\frac{i}{4\pi^{2}}\int_{\Sg} Tr\left(  \f F_{A}  + \frac{1}{2} \p
\p  \right) \, , \label{sbf}
\ee
which is a simple generalization of (\ref{bf}). Note that the fields
$\p$ have no dynamics at all. Supersymmetry (\ref{ttr}) fixes the
relative coefficient of the two terms. Just as for the action (\ref{bf})
there will be $\f$ zero modes. One `thickens' this action out as well to
\be
L = \frac{i}{4\pi^{2}}\int_{\Sg} Tr\left(  \f F_{A}  + \frac{1}{2} \p
\p  \right) + \frac{\ep}{8\pi^{2}}\int_{\Sg} Tr \f * \f  \, , \label{sbf1}
\ee
with corresponding path integral
\be
Z_{\Sg}(\ep) =\int DA D\f \exp{\left(\frac{i}{4\pi^{2}}\int_{\Sg} Tr\left(
\f  F_{A}  + \frac{1}{2} \p
\p  \right) + \frac{\ep}{8\pi^{2}}\int_{\Sg} Tr \f * \f \right)} \, .
\label{ymp}
\ee
The exact relationship between the
theories defined by (\ref{dbf}) and (\ref{sbf1}) will be given at the end
of this section.

One of the important properties of the partition function associated
with the action (\ref{sbf}) is that there is a canonical choice of
measure. On making a choice for $DA$ we pick the same for $D\p$; this is
supersymmetry preserving and the product $DAD\p$ does not depend on the
choices made. Put another way, if we send $A \rightarrow \lambda A$ then
so as not to change the transformation rules (or the relative coefficients
in the action) we must also send $\p \rightarrow \lambda \p$, and then there
is no net effect on the measure, $DAD \p \rightarrow D(\lambda A)D(\lambda
\p)  = DA D \p$.

Let us now interpret the extra term $\frac{1}{8\pi^{2}}\int_{\Sg} Tr \p \p
$ as a symplectic form on $\A$. Recall that a symplectic form $\o$ on a
$2n$-dimensional manifold is a non-degenerate two form ($det\o \neq 0$)
which is closed ($d\o=0$). There is a natural symplectic form on $\A$
which is inherited from the two manifold $\Sg$. If $a$ and $b$ are
tangent vectors in $\A$, that is $a,b \in \Omega^{1}(\Sg, Lie G)$, then
one may construct the symplectic form
\be
\Omega(a,b) = \frac{1}{8\pi^{2}}\int_{\Sg} Tr \left(a \wedge b \right)
\, . \label{symp}
\ee
That $\Omega(\, , \,)$ is closed is obvious as it does not depend on the
point $A \in \A$ at which it is evaluated. Invertibility is also clear.
We see directly that $\frac{1}{8\pi^{2}}\int_{\Sg} Tr \left( \p \wedge \p
\right) = \Omega(\p, \p)$ represents the symplectic two form of $\A$.

For a finite dimensional symplectic manifold $M$, of dimension $2m$, an
integral analogous to (\ref{ymp})
\bea
& & \int_{M} d^{2m}x d^{2m}\p \exp{\left( \frac{1}{2}\p^{\m} \o_{\m  \n}
\p^{\n}\right)} \nonumber \\
& & = \int_{M} d^{2m}x d^{2m}\p \frac{(\frac{1}{2}\p . \o
.\p)^{m}}{m!}  \nonumber \\
& &  = \int_{M} \frac{\o^{m}}{m!} \, ,
\eea
yields the symplectic volume of $M$.

We then have the immediate consequence that the partition function of
(\ref{sbf}) (or the $e^{2} \rightarrow 0$ limit of (\ref{ymp}))
evaluates the symplectic volume of $\M$.

\noindent \underline{Observables}

There are three `obvious' conditions that an observable ${\cal O}$
(a functional of the fields) should
satisfy in a topological gauge theory. These are gauge invariance, BRST
invariance and metric independence. The third may be relaxed as we will
see later. There is still a fourth condition so as not to get trivial
observables.
This is that ${\cal O} \neq \d \Theta$ for any globally defined
$\Theta$\footnote{A more detailed account of when observables are
trivial or not may be found in \cite{ot,kanno,bbrt}.}.
For if
${\cal O} = \d \Theta $ then by (\ref{dexact}) its expectation value
vanishes. Indeed this tells us that the observables must be BRST
equivalence classes of gauge invariant and metric independent functionals
of the fields. Two observables ${\cal O}_{1}$ and ${\cal O}_{2}$ are
BRST equivalent (and have the same expectation value) if ${\cal O}_{2}
=  {\cal O}_{1} + \d \Theta$ for any globally defined $\Theta$.

On the space $\Sg \times \AG$ we have the exterior derivative $d + \d$
and the curvature form $F + \p +\f$. There is also the Bianchi identity
\be
(d_{A} + \d ) (F_{A} + \p +\f) = 0 \, ,
\ee
from which we may derive the equations
\be
(d + \d ) Tr (F_{A} + \p +\f)^{n} = 0 \, . \label{desc}
\ee
Let $n=2$ and write
\be
\frac{1}{2}Tr (F_{A} + \p +\f)^{2} = \sum_{i=0}^{4} {\cal O}_{i} \, ,
\ee
where the ${\cal O}_{i}$ are $i$-forms with Grassmann grading
$(-1)^{(4-i)}$ and are given by
\bea
& & {\cal O}_{0} = \frac{1}{2}Tr (\f \f) \, , \; \; \; \; \; \; \; \;
 \; \; \; \; \; \;  {\cal O}_{1} = Tr (\p \f) \, , \nonumber \\
& & {\cal O}_{2} = Tr (F_{A}\f + \frac{1}{2}\p \p )  \, , \; \; \; {\cal
O}_{3} = \frac{1}{2}Tr(F_{A}\p) \, , \nonumber \\
& & {\cal O}_{4} = \frac{1}{2}Tr (F_{A} F_{A}) \, .
\eea
Expand the `descent' equation (\ref{desc}) in terms of form degree and
Grassmann grading as
\bea
 \d {\cal O}_{0}& =& 0 \, , \nonumber \\
 \d {\cal O}_{1}& =& -d {\cal O}_{0} \, , \nonumber \\
 \d {\cal O}_{2}& =& -d {\cal O}_{1}  \, , \nonumber \\
 \d {\cal O}_{3}& =& - d {\cal O}_{2} \, , \nonumber \\
 \d {\cal O}_{4}& =& -d  {\cal O}_{3} \, , \nonumber \\
               0& =& -d {\cal O}_{4} \, . \label{desc1}
\eea
The ${\cal O}_{i}$ are clearly gauge invariant and metric independent.
The basic observables in
the non-Abelian models on $\Sg$ are built from the ${\cal O}_{i}$ for $i=0,1
,2$. The first of these is ${\cal
O}_{0}(x) = \frac{1}{2}Tr\left( \f(x) \f(x) \right) $ which is BRST invariant,
not BRST exact, but appears to
depend on the point $x$ at which it is evaluated. Within the path integral
this is not the case,
\be
d \int_{\Phi} e^{L} \, Tr (\f \f)(x)/2 = - \int_{\Phi} e^{L} \d {\cal
O}_{1} = 0 \, .
\ee
Likewise integrating ${\cal O}_{1}$ over a one cycle $\gamma$ gives a
BRST invariant observable
\be
{\cal O}_{1}(\gamma) = \int_{\gamma} Tr ( \p \f) \, ,
\ee
\be
 \d \int_{\gamma} Tr ( \p \f) =
- \int_{\gamma}d {\cal O}_{0} = 0 \, .
\ee
That the expectation value of $\int_{\gamma} Tr ( \p \f)$ depends
only on the homotopy class of $\gamma$ may be seen as follows. Add to
$\gamma$ a homotopically trivial piece $\d \gamma = \partial \Gamma$, then
\bea
& & \int_{\gamma + \d \gamma} Tr ( \p \f) - \int_{\gamma} Tr ( \p \f)
\nonumber \\
&=&
\int_{\d \gamma} Tr ( \p \f) \nonumber \\
& =& \int_{\Gamma } dTr ( \p \f) \nonumber \\
&=& -\d  \int_{\Gamma} {\cal O}_{2} \, .
\eea

The third observable is the integral of ${\cal O}_{2}$ over the Riemann
surface,
\be
\int_{\Sg} Tr (F_{A}\f + \frac{1}{2}\p \p ) \, ,
\ee
with BRST invariance established as for the other observables.

The last observable is the supersymmetric action (\ref{sbf}) itself!
Indeed taking into account that $\frac{1}{2}Tr (\f \f)$ is essentially
independent of the point where it is evaluated the thickening term in
(\ref{sbf1}) is also essentially an observable,
\be
\int_{\Sg} \frac{1}{2}Tr (\f *\f) \sim  {\cal O}_{0} \, .
\ee

\noindent \underline{Observables In Terms Of The Partition Function}

We now show how the expectation values of the observables are determined
from the partition function
\be
Z_{\Sg}( \ep) = \int DA \, D\p \, D\f \exp{ \left(
\frac{i}{4\pi^{2}} \right.  \int_{\Sg} Tr \left(  \f F  + \frac{1}{2} \p
\p  \right) + \left. \frac{\ep}{8\pi^{2}}\int_{\Sg} Tr \f * \f
\right)} \, .
\ee

The first example is afforded by considering powers of ${\cal O}_{0}$,
\bea
<\prod_{i=1}^{k}\frac{1}{4\pi^{2}}{\cal O}_{0}(x_{i})>_{\ep}
& =& \int DA \, D\p \, D\f \exp{\left(
\frac{i}{4\pi^{2}} \right.  \int_{\Sg} Tr\left(  \f F  + \frac{1}{2} \p
\p  \right) } \nonumber \\
&  &  \; \; \; \; \; \; \; \left.  + \frac{ \ep }{8\pi^{2}} \int_{\Sg} Tr \f *
\f
\right) . \prod_{i=1}^{k} \frac{1}{8\pi^{2}} Tr \f^{2}(x_{i})
  \, . \label{ob1}
\eea
In the path integral the position of $Tr \f^{2}(x_{k})$ is immaterial so
we may replace this with $\int_{\Sg} Tr \f * \f$, using the measure with
unit area. We find
\bea
<\prod_{i=1}^{k}\frac{1}{4\pi^{2}}{\cal O}_{0}(x_{i})>_{\ep}
& = & \int DA \, D\p \, D\f \exp{\left(
\frac{i}{4\pi^{2}} \right.  \int_{\Sg} Tr\left(  \f F  + \frac{1}{2} \p
\p  \right) } \nonumber \\
& &  \; \; \; \; \; \; \; \; \; \left. +  \frac{\ep}{8\pi^{2}}\int_{\Sg} Tr \f
* \f
 \right)  . \left( \frac{1}{8\pi^{2}} \int_{\Sg} Tr \f * \f \right) ^{k} \,
, \label{ob1a}
\eea
which is
\be
\frac{\partial^{k} Z_{\Sg}(\ep)}{\partial \ep^{k}} \, .
\ee

As a second example consider
\bea
< \prod_{i=1}^{n}\frac{1}{4\pi^{2}}{\cal O}_{1}(\gamma_{i})>_{\ep}
 & = & \int DA \, D\p \, D\f \exp{\left(
\frac{i}{4\pi^{2}} \right.  \int_{\Sg} Tr\left(  \f F  + \frac{1}{2} \p
\p  \right) } \nonumber \\
& & \; \; \; \; \; \; \; \; \; + \left. \frac{\ep}{8\pi^{2}}\int_{\Sg} Tr \f
*  \f
\right) . \prod_{i=1}^{n} \frac{1}{4\pi^{2}} \oint_{\gamma_{i}} Tr \f \p
  \, . \label{int}
\eea
Here $n$ must be even or this vanishes. The action is invariant under $\p
\rightarrow - \p$ while the integrand changes sign if $n$ is odd. A simple
way to perform this integral is to introduce $n$ anti-commuting
variables $\e_{i}$ and consider instead the partition function
\bea
Z_{\Sg}(\ep, \e_{i})& =& \int DA \, D\p \, D\f \, \exp{\left(
\frac{i}{4\pi^{2}} \right.  \int_{\Sg} Tr\left(  \f F  + \frac{1}{2} \p
\p  \right)} \nonumber \\
& & + \left. \frac{\ep}{8\pi^{2}}\int_{\Sg} Tr \f * \f +
\frac{1}{4\pi^{2}} \sum_{i=1}^{n}  \e_{i} \oint_{\gamma_{i}} Tr \f \p
\right) \, . \label{int1}
\eea
On differentiating this with respect to each of the $\e_{i}$ (in the
order $i=n$ to $i=1$) and then setting these Grassmann variables to zero
one obtains (\ref{int}). Now we introduce De Rham currents $J$ with the
following properties
\be
\int_{\Sg} J(\gamma_{i}) \Lambda = \oint_{\gamma_{i}} \Lambda \, , \;
\; \; dJ = 0 \,
\ee
for any one form $\Lambda$. One completes the square in (\ref{int1}) in
the $\p$ field
\be
\p \rightarrow \p - i\sum_{i=1}^{n} \e_{i}J(\gamma_{i}) \f \, ,
\ee
to obtain
\bea
Z_{\Sg}(\ep, \e_{i})& =& \int DA \, D\p \, D\f \exp{\left(
\frac{i}{4\pi^{2}} \right.  \int_{\Sg} Tr\left(  \f F  + \frac{1}{2} \p
\p  \right) } \nonumber \\
& & + \left. \frac{\ep}{8\pi^{2}}\int_{\Sg} Tr \f * \f -
\frac{i}{4\pi^{2}} \sum_{i<j}^{n}  \e_{i} \e_{j}\int_{\Sg}
J(\gamma_{i})J(\gamma_{j}) Tr \f \f \right) \, . \label{int2}
\eea
The terms with $i=j$ vanish as $\e_{i}^{2}=0$, so that there are no
problems with self intersections.
The De Rham currents have delta function support onto their associated
cycles so that, for any zero form $\Psi$, ($i \neq j$)
\be
\int_{\Sg} J(\gamma_{i})J(\gamma_{j}) \Psi = \sum_{P \in \gamma_{i}
\cap \gamma_{j}} \sigma(P) \Psi(P) \, ,
\ee
with $P$ the points of intersection of $\gamma_{i}$ and $\gamma_{j}$
and $\sigma(P)$ ($= \pm 1$) the oriented intersection number of
$\gamma_{i}$ and $\gamma_{j}$ at $P$. This means that
\bea
& & \frac{1}{4\pi^{2}} \sum_{i<j}^{n}  \e_{i} \e_{j}\int_{\Sg}
J(\gamma_{i})J(\gamma_{j}) Tr \f \f \nonumber \\
& =& \frac{1}{4\pi^{2}} \sum_{i<j}^{n}
\e_{i} \e_{j} \gamma_{ij}  Tr\f^{2}(P)
\nonumber \\
& =& \frac{1}{4\pi^{2}} \sum_{i<j}^{n}  \e_{i} \e_{j} \gamma_{ij}
\int_{\Sg} Tr \f *\f \, ,
\eea
where we have used the fact that $Tr\f^{2}$ does not depend on the point
at which it is evaluated and $\gamma_{ij}= \# (\gamma_{i} \cap \gamma_{j})$
is the matrix of oriented intersection numbers. Putting all the pieces together
we arrive at
\be
Z_{\Sg}(\ep,\e_{i}) = Z_{\Sg}(\hat{\ep}) \, ,
\ee
with
\be
\hat{\ep} = \ep - 2 \sum_{i<j} \e_{i}\e_{j}\gamma_{ij} \, .
\ee

For $n=2$ we obtain
\bea
< \frac{1}{4\pi^{2}}{\cal O}_{1}(\gamma_{1})  \frac{1}{4\pi^{2}}{\cal
O}_{1}(\gamma_{2}) >_{\ep} & =& \frac{\partial}{\partial
\e_{1}}\frac{\partial}{ \partial \e_{2}} Z_{\Sg}(\ep
-2\e_{1} \e_{2} \gamma_{12}) \nonumber \\
&=& 2i \gamma_{12} \frac{\partial}{\partial \ep } Z_{\Sg}(\ep) \, . \label{ob2}
\eea
Likewise for higher values of $n$ the expectation values of
$\frac{1}{4\pi^{2}} {\cal O}_{1}(\gamma_{i}) $ are obtained on
differentiating $Z_{\Sg}(\ep)$.

Clearly, expectation values of mixed products
\be
< \prod_{i=1}^{k}\frac{1}{4\pi^{2}}{\cal O}_{0}(x_{i}) \prod_{j=1}^{n
}\frac{1}{4\pi^{2}}{\cal O}_{j}(\gamma_{1}) >_{\ep} \, , \label{genobs}
\ee
are similarly obtained.

\noindent \underline{Integration On Moduli Space}

We have the observables and a way of computing them, at least in
principle, but what is lacking, however, is their interpretation. We should
think of the $\p$'s as one-forms on $\AG$ and the $\f$'s as two-forms on
$\AG$. When $\AG$ is restricted to $\M$, $\p$ and $\f$ should be thought
of as a one-form and a two-form on the moduli space respectively. This
means that ${\cal O}_{0}$ is a four-form on $\M$ while ${\cal O}_{1}$ is a
three-form there.

On any $n$ dimensional manifold we may integrate an
$n$-form without the need to introduce a metric. The moduli space has
dimension (for $g>1$) $(2g-2) dim\,G$ so that any product of the
observables as in (\ref{genobs}) with $4k + 3n = (2g-2) dim\,G$ is a
form that may be integrated on $\M$. On the other hand, once the
constraint that $F_{A}=0$ has been imposed, the path integral over $\AG$
devolves to an integral over $\M$. In this way (\ref{genobs}) is seen to
be the integral over $\M$ of a $(2g-2) dim\,G$-form. Let us denote with
a hat the differential form that an observable corresponds to. Then
(\ref{genobs}) takes the more suggestive form
\be
< \prod_{i=1}^{k}\frac{1}{4\pi^{2}}{\cal O}_{0}(x_{i}) \prod_{j=1}^{n
}\frac{1}{4\pi^{2}}{\cal O}_{j}(\gamma_{1}) >_{\ep} = \int_{{\cal M}_{F}}
\prod_{i=1}^{k}\frac{1}{4\pi^{2}}\hat{{\cal O}}_{0}(x_{i}) \prod_{j=1}^{n
}\frac{1}{4\pi^{2}}\hat{{\cal O}}_{j}(\gamma_{1}) \exp{\Omega} \, .
\ee
When $4k +3n = (2g-2) dim\,G$, the symplectic form makes no
contribution. However, if $4k +3n = 2m < (2g-2) dim\,G$ there will also
be contributions from the action to soak up the excess form-degree.
On expanding the exponential, the symplectic form $\Omega(\p,\p)$ raised to
the power $(g-1)dim\,G -m$ will survive the Grassmann integration.

We now know that we are calculating integrals of metric independent
differential forms on $\M$. What do such integrals correspond to? They
are naturally interpreted as intersection numbers on $\M$. An
explanation of the relationship between the differential form and
intersection viewpoints is provided in \cite{bbrt}.

\noindent \underline{Relationship Between The Old And The New}

The problem with the presence of $B$,
$\chi$, $\bar{\f}$ and $\eta$ zero modes is that the theory, as it stands,
is not defined. It is possible, however, to deal directly with these
modes. We add to the action a term that is
supersymmetric and that damps them. The new action is
\bea
L(t) &=& L + t \d \, i\int_{\Sg} Tr \chi *\bar{\f} \nonumber \\
     &=& i \int_{\Sg} Tr \left( BF_{A} - \chi d_{A} \p + \eta
d_{A} * \p  + \bar{\f} d_{A}*d_{A} \f + \bar{\f} \{ \p, * \p \} \right)
\nonumber \\
& & \; \; \; \; \; +  t\, i \int_{\Sg} Tr \left(B * \bar{\f} - \chi * \eta
\right) \,
{}.
\eea
By an argument that is similar to the proof of metric independence
(\ref{metv1})-(\ref{dexact}), the path integral defined by this action
is independent of smooth variations of $t$. One must be careful,
however, as the $t\rightarrow 0$ limit is not the same as taking $t=0$
directly precisely because of the presence of zero modes.

Integrating out the fields $B$, $\bar{\f}$, $\c$ and $\e$ generates a new
action solely in terms of the geometric fields,
\be
L'(t) = \frac{i}{t} \int_{\Sg} Tr \left(*F_{A}d_{A}*d_{A}\f + *d_{A} * \p
d_{A} \p + F_{A} \{ \p , *\p \} \right) \, .
\ee
The integral over $\f$ lands us on the space of solutions to
$d_{A}*d_{A}*F_{A}=0$, but this equation is the same as the Yang-Mills
equation $d_{A}*F_{A}=0$. One proves this by considering
\bea
0& =& \int_{\Sg} Tr F_{A}* d_{A}*d_{A}*F_{A} \nonumber \\
&=&  \int_{\Sg}Tr (d_{A}*F_{A})*(d_{A}*F_{A}) \, ,
\eea
the last line being the norm of $d_{A}*F_{A}$, which vanishes. Hence
$d_{A}*F_{A} =0$. In passing to the new action we have moved away from
just a description of the flat connections, and find that all the
Yang-Mills connections contribute.

Now suppose that we wish to compute the expectation value in this new
theory of
\be
\exp{\left(
\frac{i}{4\pi^{2}}   \int_{\Sg} Tr\big(  \f F  + \frac{1}{2} \p
\p \big)  +  \frac{\ep}{8\pi^{2}}\int_{\Sg} Tr \f * \f  \right)} \, .
\label{expv}
\ee
The expectation value continues to be independent of $t$. We may,
therefore, set $t = \infty$ as the theory remains well defined for this
value. This is the correspondence we were looking for. The partition
function of Yang-Mills theory that we have been using (\ref{ymp}) is, in
terms of the original model, the expectation value of (\ref{expv}).
The expectation values of the observables (\ref{ob1})-(\ref{ob2}) are
the same in the original theory as long as it is understood that
(\ref{expv}) is inserted.

\setcounter{equation}{0}
\section{ U(1) Theory}

\noindent \underline{First Chern Class}

$U(1)$ bundles over a Riemann surface $\S _{g}$ are classified by their
first Chern class
\be
c_{1} = \frac{1}{2\pi} \int_{\Sg} F_{A} \, , \label{cclass}
\ee
which is an integer, say $k$, and for the Abelian theory $F_{A}=dA$. The
most familiar configuration that
has a non vanishing first Chern class is the  magnetic monopole of
Dirac. On $S^{2}$, for example, we may consider a connection $A_{+}$ on
the northern hemisphere $H_{n}$ and $A_{-}$ on the southern $H_{s}$. These
`patch' together, if on the equator, where they overlap, they agree up to
a gauge transformation. This means that on the equator there exists a
$\vf$ such that
\be
A_{+} = A_{-} + d\vf \, .
\ee
The Chern class may be expressed as
\bea
\frac{1}{2\pi} \int_{S^{2}} F_{A}& =& \frac{1}{2\pi} \int_{H_{n}}dA_{+} +
\int_{H_{s }}dA_{-} \nonumber \\
&=& \frac{1}{2\pi}\oint A_{+} - \frac{1}{2\pi} \oint A_{-} \nonumber \\
&=& \frac{1}{2\pi}\oint d \vf \, ,
\eea
with the relative sign appearing due to the opposite orientation of the
circle boundary of the northern and southern hemispheres. When, in local
coordinates, $\vf = k \t$ the first Chern class is $k$. Such a $\vf$
is allowed as it corresponds to a periodic group element, $\exp{(k i
\t)}$, with $ 0 \leq \t <2\pi$.

On the Torus $T^{2}= S^{1}\times S^{1}$ with local (angular) coordinates
$(\sigma_{1},\sigma_{2})$ the gauge field
\be
A = \frac{m}{2\pi} \sigma_{1}d\sigma_{2} + \frac{n}{2\pi}
\sigma_{2}d\sigma_{1 } \, , \label{ct}
\ee
satisfies
\bea
k & =&  \frac{1}{2\pi}\int_{T^{2}} F_{A} \nonumber \\
  &=& \frac{1}{4\pi^{2}} \int_{T^{2}} (m-n) d \sigma_{1} d \sigma_{2}
\nonumber \\
  &=& m-n \, .
\eea
The gauge field is not periodic, but it is periodic up to a gauge
transformation. If we send $\sigma_{1} \rightarrow \sigma_{1} + 2\pi$
then
\be
A(\sigma_{1}+2\pi,\sigma_{2}) = A(\sigma_{1},\sigma_{2}) +
e^{im\sigma_{2}}id e^{-im\sigma_{2}} \, ,
\ee
with a similar relationship for $\sigma_{2} \rightarrow \sigma_{2} +
2\pi$. The gauge group elements are globally defined.

\noindent \underline{Flat Connections}

We have seen that the moduli space of flat $U(1)$ connections on a genus
$g$ surface, ${\cal M}_{F}(\S_{g}, U(1) )$, has dimension $2g$. Indeed it is a
$2g$ torus ${\cal M}_{F}(\S _{g}, U(1) ) = T^{2g}$. Let us see how it comes
out for genus zero and one.

For the sphere ($g=0$) all loops are contractible and the only flat
connection, up to gauge equivalence, is the trivial connection. The
moduli space is therefore a point. For the torus ($g=1$) there are two
possible non-trivial holonomies. The corresponding flat gauge field has
the form
\be
A = \frac{\alpha_{1}}{2\pi} d\sigma_{1} +\frac{\alpha_{2}}{2\pi}  d \sigma _{2}
\, .
\ee
But what are the ranges of $\alpha_{1}$ and $\alpha_{2}$? Note that we may
still perform (single valued) gauge transformations
\be
A \rightarrow A + e^{(i m_{1} \sigma_{1} + i  m_{2}
\sigma_{2}) }i d e^{(-i m_{1} \sigma_{1} - i m_{2}
\sigma_{2}) } \, ,
\ee
which corresponds to the shifts
\be
\alpha_{i} \rightarrow \alpha_{i} + 2\pi m_{i} \, ,
\ee
for all integers $m_{i}$. In other words the gauge inequivalent $A$ have
$\alpha_{i}$ that live on $T^{2}$. This then is the space ${\cal M}_{F}(T^{2},
U(1) )$.

This correspondence between the holonomies of the flat gauge fields and
the points of ${\cal M}_{F}(\Sg, U(1) )$ can be made even more
explicit. The local coordinates of ${\cal M}_{F}(\Sg, U(1) ) =
T^{2g}$ are simply
\be
(\oint_{a_{1}}A, \dots ,\oint_{a_{g}}A, \oint_{b_{1}}A, \dots
,\oint_{b_{g}} A) \, .
\ee

In order to establish that the moduli space is a torus, in general, we would
need some more notions from the theory of Riemann surfaces, so we forgo this.

\noindent \underline{Maxwell Connections}

These are defined to be the class of connections that satisfy the
Maxwell equation
\be
d*F_{A} = 0 \, .
\ee
In terms of the zero-form, $f_{A}=*F_{A}$, this equation becomes
\be
df_{A}= 0 \, ,
\ee
which has as its solutions the harmonic functions $f_{A} \in
H^{0}(M,R)$. On a compact manifold these are the constant functions, so
that we find
\be
F_{A} = a \omega \, ,
\ee
where $a$ is some constant and $\omega$ is a volume form normalised to
unity. On a bundle with first Chern class equal to $k$ we have
\be
F_{A}= 2\pi k \omega \, . \label{maxk}
\ee
This last equation is equivalent to the original Maxwell equation, and
when $k=0$, the Maxwell connections are flat. The
connection (\ref{ct}) on $T^{2}$ is a Maxwell connection.

The moduli space of Maxwell connections, ${\cal M}_{M}^{k}(\Sg , U(1))$ is
the same as the moduli space of flat connections, that is
\be
{\cal M}_{M}^{k}(\Sg , U(1)) = {\cal M}_{F}(\S _{g}, U(1) ) \, ,
\ee
and as this correspondence holds for any $k$, we may supress it. To see that
this must be
true, let $A_{k}$ be any Maxwell connection satisfying (\ref{maxk}).
Then all other connections on the ($c_{1}=k$) bundle are of the form
\be
A=A_{k} + X \, , \label{dec}
\ee
for some one from $X$. For $A$ to also be a Maxwell
connection $X$ must satisfy
\be
dX =0 \, , \label{har}
\ee
which is the flatness equation and does not depend on $k$. Gauge
inequivalent $X$'s are the points of the moduli space ${\cal M}_{M}(\Sg ,
U(1)) $, but clearly are also the points of ${\cal M}_{F}(\Sg , U(1))$.

There is a more geometric way of stating this. We have seen that, for
flat connections, we can form a map from $\pi_{1}(M)$ to $G$ and,
conversely, that these maps, up to conjugation, characterise the moduli
space of flat connections. We can likewise show, that given Maxwell
connections on any surface $\Sg$, we may form maps from $\pi_{1}(\Sg)$
to $G$. Fix a Maxwell connection $A_{k}$. Then for any Maxwell
connection $A$, also with first Chern class equal to $k$, we can form
the required map $\hat{\vf}_{\g}(A) = \vf_{\g}(A_{k})^{-1}
\vf_{\g}(A)$. We see that $\hat{\vf}_{\g}(A)$ depends only on the
homotopy class of $\g$, for varying the path we get an area contribution
from $\vf_{\g}(A_{k})^{-1}$ that cancels that from $\vf_{\g}(A)$ (the area
dependence may be seen in the second last line of (2.8)).

\subsection{Maxwell theory on compact closed surfaces}

We take the classical action of Maxwell theory on a two-dimensional
(orientable) surface to be
\be
L=\frac{1}{4\pi^{2}}\int_{\Sg} i \f F_{A}  - \frac{\ep}{8\pi^{2}}
\int_{\Sg } \f * \f  \, . \label{1}
\ee

The partition function of Maxwell theory in the topological
sector with monopole charge (first Chern class) $k$,
\[\frac{1}{2\pi}\int_{\Sg}F_{A} =  k\;\;,\]
is then
\be
Z_{\Sg}(k,\ep) = \int DA \, D\f \, \exp{\left(\frac{1}{4\pi^{2}}\int_{\Sg} i \f
F_{A}  - \frac{\ep}{8\pi^{2}}
\int_{\Sg } \f * \f  \right)} \d\big(\frac{1}{2\pi}\int_{\Sg}F_{A} -
k\big)\;\; .
\label{3}
\ee
This still needs to be gauge fixed, but we make use of the trivialising
map so we forgo the introduction of the fields associated with the gauge
fixing proceedure and pass directly to the partition function in the form
\bea
Z_{\Sg}(k,\ep)& =& \int D\xi D\f \, \exp{\left(\frac{1}{4\pi^{2}}\int_{\Sg} i
\f
\xi  - \frac{\ep}{8\pi^{2}}
\int_{\Sg } \f * \f  \right)} \d\big(\frac{1}{2\pi}\int_{\Sg}\xi -  k\big)
\nonumber \\
 &=& \int D\xi \, \exp{\left(\frac{-1}{8\pi^{2}\ep}\int_{\Sg}  \xi *
\xi  \right)} \d\big(\frac{1}{2\pi}\int_{\Sg}\xi -  k\big)\, . \label{4}
\eea

Introducing a multiplier $\l$ to represent the delta-function as
\[\d(\frac{1}{2\pi}\int_{\Sg} \xi -  k) =  \int_{-\infty}^{+\infty} d\l \,
\exp{\left(i \l (\int_{\Sg}\xi - 2\pi k) \right)}\;\;,\]
the Gaussian integrals over $\xi$ and $\l$ are easily performed to give
\be
Z_{\Sg}(k,\ep) =
   \frac{1}{\sqrt{ 2\pi \ep}}\exp{\left(\frac{-k^{2}}{2\ep } \right)}
   \label{5}\;\;.
\ee
Note that $Z_{\Sg}(k,\ep)$ is independent of the genus of $\Sg$.

\noindent \underline{Fixed point theorems}

Apart from the universal factor $1/\sqrt{2\pi \ep}$, which arises
from the reducibility of the connections, the partition function
(\ref{3}) is given entirely by the contribution at the Maxwell connection
(\ref{maxk}). That is, (\ref{5}) may be re-written as
\be
Z_{\Sg}(k,\ep) =
   \frac{1}{\sqrt{2\pi \ep}}\exp{\left( L(A_{k}) \right)} \, .
\ee
Furthermore, if we sum over the different topological sectors to calculate the
overall partition function we find
\bea
Z_{\Sg}(\ep) &=& \sum_{k} Z_{\Sg }(k,\ep) \nonumber \\
        & = & \sum_{k} \frac{1}{\sqrt{2\pi \ep}} \exp{\left( L(A_{k}) \right)}
\, . \label{maxdh}
\eea
These results are rather astounding. They tell us that the entire
contribution to the path integral comes simply from the values at the
critical points of the action. The critical points
being the Maxwell connections.

Perhaps the importance of the result is
overshadowed by its `obviousness'. We have only had to perform Gaussian
integrals and these {\em are} evaluated by their equations of motion;
in (\ref{gaus}) this is $x =y/e^{2}$, which, when substituted back into
the `action' yields the exponent on the right hand side. But two facts
conspired to turn the problem into one of simple Gaussian integration.
The availability of a trivialising map and the fact that the Maxwell
equations become `algebraic' (\ref{maxk}). The conspiracy continues
unabated in the non-Abelian theory \cite{w2dr}.

For integrals over finite dimensional
symplectic manifolds such reductions to the fixed point set of the
action (exponent) are explained in terms of the fixed point theorems of
Duistermaat and Heckman \cite{dh}. Witten has generalised these theorems to the
non-Abelian case and an infinite dimensional setting (the manifold
$\A$). Quantum Maxwell theory furnishes a very simple example of these ideas.

\noindent \underline{Reducible Connections And $vol\, {\cal M}(\Sg, U(1) )$}

For the Abelian theory, reducible connections do not constitute a real
problem as all Abelian connections are reducible and in the same way. This
means that we may extract an overall contribution from the
constant $\f$ field,
\be
\int_{-\infty}^{+\infty} d\vf
\exp{\left(-\frac{\ep}{8\pi^{2}}
 \vf^{2}\right)} = \frac{2\pi\sqrt{2\pi}}{\sqrt{\ep}} \, .
\ee
Dividing this out of the partition function $Z_{\Sg}(k,\ep)$ gives
\be
\hat{Z}_{\Sg}(k,\ep) = \frac{1}{4\pi^{2}}
\exp{\left( \frac{-k^{2}}{2\ep}\right)} \, .
\ee
Clearly as $\ep \rightarrow 0$ this vanishes for all $k$ except $k=0$.
This is consistent with the fact that we should land on the flat
connections in the limit.

For $k=0$ at $\ep=0$ we have, tentatively,
\be
\hat{Z}_{\Sg} = vol \, {\cal M}_{F} = \frac{1}{4\pi^{2}} \, .
\ee
This result should not be taken too seriously, as there are many factors
that we have not been able to fix uniquely (such as the normalization of
the path integral measure). These factors, however, will not be $k$
dependent, and they will have a smooth $\ep \rightarrow 0$ limit.
Nevertheless we see that it is possible to
obtain a finite expression, and, in principle, with a more careful
analysis, a correct form for $vol{\cal M}_{F}$.

\noindent \underline{Non Contribution Of Harmonic One-Forms }

The reason for needing more care is that in one sense we have missed the
volume we are looking for altogether! The trivializing map is invertible
everywhere in field space (that is in $\AG$) outside a finite
dimensional set, points of which are in one to one correspondence with
the space of flat connections. These are the fields $X$ in (\ref{dec})
which satisfy (\ref{har}) and are gauge fixed $d*X=0$, so that they are
harmonic one-forms. The partition function (\ref{5}) is then
still to be multiplied by the $vol\,{\cal M}_{F} $. The result for the
partition function, up to a standard renormalization (see section
\ref{pfym}), that we have derived, is re-obtained in the next section.
The standard renormalization implies, that for genus $g$, the partition
function has the form
\be
Z_{\Sg}(k,\ep) =
   \frac{\kappa^{(2-2g)}}{\sqrt{ 2\pi \ep}}\exp{\left(\frac{-k^{2}}{2\ep
 }  \right)} \, ,
\ee
for some $\kappa$. Clearly a different input is required to fix the
constant.

\noindent \underline{Triviality Of Wilson Loops Along Homology Cycles}

The harmonic forms also do not contribute to correlation functions of operators
which can be expressed in terms of $\xi$ $(=F_{A})$, and the volume
of the moduli space of flat connections will consequently drop out of
normalized correlation functions in this case as well.
One may wonder, however, what happens to correlation functions of operators
which are sensitive to the holonomies of the gauge fields along the homology
cycles of $\S_{g}$. The gauge invariant observables of interest to us here are
Wilson loops
\[\exp{\left(i\a\oint_{\g}A\right)}\]
along closed loops $\g$. If $\g$ is homologically trivial
then - by Stokes theorem -
the Wilson loop is expressible in terms of $\xi$ and thus falls into the
category of operators already dealt with above. One may have some doubts on
the validity of Stoke's theorem for connections on non-trivial bundles
($k\neq 0$) but for $k\a\in{\bf Z}$ Stoke's theorem can indeed still be used
in the exponent. This is precisely analogous to the quantization condition in
the WZW action, and we will derive this condition below.

This leaves us with Wilson loops for homologically non-trivial $\g$,
which are indeed sensitive to the holonomies of $A$. In this case $\a$ has
to be an integer in order to define a gauge invariant operator (under
the large gauge transformations). With $\a\in{\bf Z}$, however, we find that
the expectation value $\langle\exp{\left(i\a\oint_{\g}A\right)}\rangle$,
as well as any correlator involving homologically non-trivial loops, is
identically zero,
\[ \langle \exp{\left(i\a\oint_{\g}A\right)}\rangle_{k} = 0\;\;\;\;\;\;\;\;\;
             \rm{for}\;\;\g\neq\partial \Gamma\;\; , \a \neq 0 . \label{twl}\]
Thus the failure of the trivializing map in this case causes no distress. One
way of proving the vanishing of this expectation value is to note the
fact that the evaluation of the holonomy of one of the $X$ is
\[\int dX \exp{\left( i\a\oint_{\g}X \right)} \sim \int_{0}^{1}d\theta
e^{2\pi  i\a\theta}= 0 \;\;\;\;\;\;\;\;\;\a\in{\bf Z},
                                                            \;\;\a\neq 0
\, , \]
as the moduli space is a torus.

Thus in the Abelian case this rules out homologically non-trivial Wilson loops
as interesting observables on closed surfaces.


\noindent \underline{A Quantization Condition And Contractible Loops}

We now turn to the computation
of correlators of any number of (possibly intersecting and self-intersecting)
contractible Wilson loops, starting with the case of a single non-intersecting
loop. The first thing to note is that on a closed surface there is an intrinsic
ambiguity in trying to write $\oint_{\g}A = \int_{D}\xi$, where $A$ is a
connection on a non-trivial bundle and
$\partial D = \g$, as one could equally well replace $D$ by its complement
$\S_{g}\setminus D = -D'$. Making a particular choice now, we will have
to inquire at the end under which circumstances
the result is independent of any such choice (and this will, as expected, give
rise to the quantization condition on $\a$). Using the same representation
for the delta function as above and performing the Gaussian integrals one
finds that the normalized expectation value is
\bea
& & \langle \exp{\left(i\a\int_{D}\xi\right)}\rangle_{k} \nonumber \\
&\equiv&
\frac{1}{Z_{\Sg}(k,\ep)}\int D\xi \, \exp{\left(-\frac{1}{8\pi^{2}\ep
}\int_{\Sg}\xi*\xi +i\a\int_{D}\xi \right)}
\d(\frac{1}{2\pi}\int_{\Sg}\xi- k) \nonumber\\
&=&\exp{\left(-2\pi^{2} \ep
\a^{2}\frac{A(D)}{A(\Sg)}\frac{A(D')}{A(\Sg)} \right)} \exp{\left(2\pi i k \a
 \frac{A(D)}{A(\Sg)} \right)}\;\;.
\label{7}
\eea
The first term is manifestly symmetric in $D$ and $D'$ and if we compute
instead $\langle\exp{\left(-i\a\int_{D'}\xi\right)}\rangle_{k}$ we find that
\[\langle \exp{\left(-i\a\int_{D'}\xi\right)}\rangle_{k}=
  \langle \exp{\left(i\a\int_{D}\xi\right)}\rangle_{k} \, \exp{\left(-2\pi
ik\a \right)}\;\;,\]
so that $\langle\exp{\left(i\a\oint_{\g}A\right)}\rangle_{k}$ can only be
defined consistently if $k=0$ or $\a=\frac{n}{k},\;\;n\in{\bf Z}$. In the first
case
the structure group of the $U(1)$ bundle can be extended to ${\bf R}$
and $\a\in{\bf R}$ labels the unitary representations of the universal
covering group ${\bf R}$ of $U(1)$. In the second case $\a$ defines a
representation of a $k$-fold covering of $U(1)$. The quantization condition
on $\a$ has, like that of the WZW model, a natural group theoretic and
geometric explanation.

When considering loops with self-intersections or several intersecting loops
no substantially new features arise and the calculations can be done in much
the same way as that leading to (\ref{7}). The result is that in the general
formula for the correlator of $n$ intersecting but non-selfintersecting loops
the exponent in (\ref{7}) is replaced by
\bea
&&-2\pi^{2} \ep
\sum_{j=1}^{n}\a_{j}^{2}\frac{A(D_{j})}{A(\Sigma)} \frac{A(D_{j}')}{A(\Sigma)}
  +2\pi ik\a\sum_{j=1}^{n}\a_{j}\frac{A(D_{j})}{A(\Sigma)}\nonumber\\
&&+4\pi^{2}\ep
\sum_{j=1}^{n} \sum_{i<j}\a_{j}\a_{i} \left(\frac{A(D_{j})}{A(\Sigma)}
\frac{A(D_{i})}{A(\Sigma)}  -A(D_{j}\cap D_{i})\right)\;\;.\label{8}
\eea
Again it can be checked (with a little bit of algebra) that the result is
independent of the choice of $D_{j}$ or $D_{j}'$ provided that
$k\a_{j}\in {\bf Z}$. Moreover, by regarding a self-intersecting loop as two
touching but non-selfintersecting loops with opposite orientations, equation
(\ref{8}) gives the general result for the correlator of intersecting and
self-intersecting loops on a closed surface of any genus.

One curious observation is useful to keep in mind when checking if the
result (\ref{8}) is sensible and correct. In flat space the figure eight loop
and the figure eight folded into itself give different results \cite{bralic}.
This is of course perfectly reasonable as the folding leads to points in the
interior part of the loop being surrounded twice by the loop so that (energy
being proportional to the flux squared) these points contribute to the path
integral with the four-fold weight of those surrounded by just one loop
(something that is also reflected in the quadratic composition law of
(\ref{8})). On the two-sphere, however, these two configurations are
indistinguishable, whereas on the torus they are again manifestly different,
and one may wonder how the path integral manages to take this into
account. As it turns out the path integral automatically gives a sensible
answer. Indeed, by staring at a figure eight on the two-sphere one can
convince oneself that the process of folding (say) the upper loop into the
lower is equivalent to going to the complement of the lower loop, and
(\ref{8}) does not depend on whether we choose one interior of a loop or
its complement.

\noindent \underline{Wilson Points}

The evaluation of expectation values of Wilson loops is also quite
straightforward. The correlator
\bea
& & \langle \exp{\left( \frac{i}{2\pi} \sum_{j}  q_{j} \f(x_{j}) \right)}
\rangle_{k} \nonumber \\
& =& \frac{1}{Z_{\Sg}(k,\ep)}\int
D\xi D\f \, \exp{\left(\frac{1}{4\pi^{2} }\int_{\Sg} i\xi \f -
\frac{\ep}{8\pi^{2} }\int_{\Sg} \f * \f  \right. } \nonumber \\
& &
  \;\;\;\;\; \; \; \; \; \; \; \left. + \; \; \; \frac{i}{2\pi} \sum_{j}
\int_{\Sg}J(x_{j})\f \right) \d(\frac{1}{2\pi}\int_{\Sg}\xi- k) \, ,
\eea
is easily evaluated by performing all the Gaussians, but may be obtained
directly by redefining $\xi$. Change variables according to ($J(x_{i})$
is a two-form De Rham current that fixes one to the point $x_{i}$)
\be
\xi \rightarrow \xi - 2\pi \sum_{j} q_{j} J(x_{j}) \, , \label{shift}
\ee
which simply acts to get rid of the Wilson points in the exponent and
shifts $k$ to $k+ \sum_{j} q_{j}$ in the delta function, so that the sum
must be an integer. We find, therefore, that
\bea
\langle \exp{\left( \frac{i}{2\pi} \sum_{j}  q_{j} \f(x_{j}) \right)}
\rangle_{k}& =& \frac{1}{\sqrt{2\pi \ep}}\exp{\left(-\frac{(k +
\sum_{j}q_{j})^{2}}{2\ep } \right)} \nonumber \\
&=& Z_{\Sg}(k+\sum_{j}q_{j},\ep) \, .
\eea

There are a few points that are worth special mention. The effect of
introducing the Wilson points is to make the theory behave as if it is
on a bundle of different Chern class. The correlation function does not
depend on the location of the Wilson points, a fact which is clear from
the cohomological field theory nature of the correlator but not so
obvious from the Yang-Mills point of view. The expectation value of
Wilson points with Wilson loops on non-trivial homology cycles will
vanish. The expectation value of Wilson points with homologically
trivial Wilson loops will reproduce (\ref{7}) but with the shift $k
\rightarrow k + \sum_{j}q_{j}$. This last result comes from the fact
that under the shift (\ref{shift})
\be
\exp{\left(i\a \int_{D}\xi \right) } \rightarrow \exp{\left(i\a
\int_{D}\xi \right) } \exp{\left(2\pi i\a \sum_{j}\int_{D} q_{j}
J(x_{j}) \right) } \, ,
\ee
and
\be
\int_{D} J(x_{j}) = (\pm 1 , 0) \, ,
\ee
depending on whether the point $x_{j} \in D$ or not. In either case the
second exponential is unity, providing we take account of the fact that
both $\a$ and $\sum_{j} q_{j}$ are integers.

\noindent \underline{Topological Observables}

As far as the Schwarz type topological observables are concerned, only
the partition function is non-trivial. We concentrate on the
cohomological observables. In the case of $U(1)$ we may also take $n=1$
in (\ref{desc}) so that the integrals of $\p$ around the homology
cycles are observables. Let us set
\be
\oint_{a_{i}} A = \a_{i} \, , \; \; \; \; \; \; \; \; \; \;
\oint_{b_{i}} A = \beta_{i} \, ,
\ee
so that the flat coordinates of the torus $T^{2g}$ are
$(\a_{i},\beta_{i})$. Then
\be
\oint_{a_{i}} \p = d\a_{i} \, , \; \; \; \; \; \; \; \; \; \;
\oint_{b_{i}} \p = d\beta_{i} \, ,
\ee
where the exterior derivative $d$ is that on $T^{2g}$ (it is $\d$
restricted to the torus). While these observables are exact, they are
not trivial as the coordinates $(\a_{i},\beta_{i})$ are not globally
defined. We have the following correspondence
\be
\langle \prod_{i=1}^{g} \oint_{a_{i}}\p \oint_{b_{i}} \p \rangle =
\int_{T^{2g}} \prod_{i=1}^{g} d\a_{i} d \beta_{i} \, .
\ee

\setcounter{equation}{0}
\section{Field Theory On Manifolds With Boundary}

The path integral on a manifold $M$, with boundary $\del M = B$, requires
boundary conditions to be fixed on $B$. In this way the path integral
becomes (for each operator insertion in the interior of $M$) a functional
of the fields on $B$, which can be regarded as a state in the
canonical Hilbert space of the theory on $B\times{\bf R}$. While this
procedure is of conceptual interest as it sets up the correspondence between
the path integral and operator formalisms of field theory, it is generally
of little practical use as the path integrals involved are too complicated
to be calculated directly. In certain cases symmetry arguments may be
invoked to determine the states uniquely (as in string theory \cite{vafa})
or up to a finite ambiguity (as in Chern-Simons theory \cite{cs}).

In the case of topological field theories it is possible to
deduce certain general properties of and relations between correlation
functions. This is in line with the axiomatic approach to topological field
theory as proposed by Atiyah \cite{at}. For two dimensional Yang-Mills
theory, which is almost topological, depending only on the measure on
the Riemann surface, and is also a gauge theory, it turns out that it is
possible to completely determine the states. In turn one may use this
information to evaluate the path integral on any surface.

We proceed to explain the underlying ideas and then we reproduce the results
we have obtained for the $U(1)$ theory using these techniques.

\noindent \underline{Boundary Data}

When given a path integral to compute on a manifold with boundary, one
must specify some boundary configuration of the fields. In equations, we have
\be
\Psi_{M}(\vf) = \int_{\f \mid_{B}= \vf} D\f \, e^{-S(\f)} \, .
\ee
Of course not all boundary data may be specified. There will, depending
on the theory at hand, be certain restrictions.

For a gauge theory, it
may be possible to demand that $\Psi_{M}$ be gauge invariant, or at
least transform in some well specified way under gauge transformations.
To see this in practice, suppose that the action $S(\f)$ is invariant under
gauge transformations $\f \rightarrow \f^{g}$ that are not the identity on
the boundary. Then we have
\bea
\Psi_{M}(\vf)& =& \int_{\f \mid_{B}= \vf} D\f \, \exp{-S(\f)} \nonumber \\
&=& \int_{\f^{g} \mid_{B}= \vf} D\f^{g} \, \exp{-S(\f^{g})} \nonumber \\
& =& \int_{\f^{g} \mid_{B}= \vf} D\f \, \exp{-S(\f)} \nonumber \\
& =& \Psi_{M}(\vf^{g^{-1}_{B}}) \, . \label{wfi}
\eea
The second equality is a change of variables of the dummy $\f$, the
third follows from the gauge invariance of the action and the presumed
gauge invariance of the path integral measure. In the fourth equality
$g_{B}$ stands for the value of the gauge parameter on the boundary.

This is the behaviour that two dimensional Yang-Mills theory exhibits.
An example where (\ref{wfi}) does not hold is Chern-Simons theory. Here
the wavefunctions pick up a phase under gauge transformations and are
properly thought of as sections of certain bundles.

For manifolds with more boundary components, the partition function is a
functional of the data on each component of the boundary.

\noindent \underline{Glueing Manifolds Together}

We want to see how to get at the partition function of a manifold by glueing
together two manifolds. For concreteness and ease of visualisation
consider the two sphere and put an imaginary line along the
equator. The path integral is an integral over all possible field
configurations on the two sphere. Pick some allowed configuration $\vf$
on the equator. We can think of performing the path integral on the two
sphere by integrating over all configurations which are consistent with
$\vf$ on the equator and then integrating over all possible $\vf$. As we
integrate over the sphere, the
path integral on the southern hemisphere gives the partition function of
the disc with boundary data $\vf$ while the path integral on the northen
hemisphere also gives the partition function of the disc (with opposite
orientation) and with boundary data $\vf$. We have deduced that
\be
Z_{S^{2}} = \int D\vf \, \Psi_{D}(\vf) \Psi_{-D}(\vf) \, .
\ee

This generalises directly to arbitrary manifolds. If $M$ is cut into two
manifolds $M_{1}$ and $M_{2}$ along $B$, then we have
\be
Z_{M} = \int D\vf \, \Psi_{M_{1}}(\vf) \Psi_{M_{2}}(\vf) \, .
\ee
Note that we do not preclude $M$ from having boundary or, equivalently,
the $M_{i}$ from having more boundary components than just $B$.

\subsection{Maxwell theory on surfaces with boundary}

\noindent \underline{The Disc}

The first thing we
have to determine is the allowed boundary conditions. If the resulting
state is to be invariant under small gauge transformations (i.e. satisfy
Gauss' law) the boundary conditions have to be chosen to be gauge invariant.
Now the only gauge invariant degree of freedom of a gauge field on the circle
$\del D$ is its holonomy $\t\in{\bf R}$ defined by
\[\oint_{\del D}A = 2\pi\t\;\;,\]
and the only admissible boundary condition is therefore the specification of
$\t$. Computing the path integral with this boundary condition then amounts
to inserting $\d(\oint_{\del D}A -2\pi\t)$ into the path integral, that
is
\bea
\Psi_{D}(\t,\ep)& =& \int_{\frac{1}{2\pi}\oint_{\del D}A = \t} \, DA \, D\f \,
 e^{L}
\nonumber \\
& \equiv & \int DA \,D\f \, e^{L} \, \, \d\big(\frac{1}{2\pi}\oint_{\del D}A
- \t\big)
\,.
\eea
This form means that we may use the trivialising map again to simplify
matters
\be
\Psi(\t,\ep) = \int D\xi D\f \, e^{L} \, \d\big(\frac{1}{2\pi} \int_{D}\xi -
\t\big) \, .
\ee

A question that arises at this point is what type of delta
function should appear here? We saw before that there are large gauge
transformations on the circle due to $\pi_{1}(U(1))={\bf Z}$. These act
on $\t$ as $\t\rightarrow\t + n,\;n\in{\bf Z}$, and we can demand invariance
under these transformations which would then render the wave function a
periodic function of $\t$. This is accomplished by inserting the periodic
delta function $\d^{P}(\int_{D}\xi -2\pi\t)$, defined by
\be
\d^{P}(x)=\sum_{n\in{\bf Z}}\d(x+2\pi n)=\sum_{n\in{\bf Z}}e^{2\pi inx}\;\;,
\label{9}
\ee
into the path integral.

With these preparatory remarks in mind we calculate, with the standard
delta function,
\be
\Psi(\t, \ep) = \frac{1}{\sqrt{2\pi\ep}}
\exp{\left(-\frac{\t^{2}}{2\ep} \right)} \, , \label{10}
\ee
and
\be
\Psi^{P}(\t, \ep) =
\sum_{n}\exp{\left(-2 \pi^{2}\ep n^{2}\right) }\exp{\left(2\pi
in\t\right)} \label{11}\;\;,
\ee
with the periodic delta function. $\Psi^{P}$ is of the form
$\sum_{n}a_{n} \c_{n}(\t)$, where $\c_{n}$ is the
character of the unitary irreducible charge $n$ representation of $U(1)$.
This is also the general form of the states of Yang-Mills theory
on the disc.

The wavefunctions (\ref{10}) and (\ref{11}) are solutions to the
heat (Schr\"{o}dinger) equation on the line and and circle repectively,
with the initial condition that they are delta functions. This is an
expected relationship between the path integral with boundary and the
Schr\"{o}dinger equation (see appendix \ref{lap}). There is also an
unexpected relationship
between (\ref{10}), (\ref{11}) and modular forms which is explained in
part in \cite{bt}.

\noindent \underline{Twisted States}

It is well known that more generally states could carry a
non-trivial unitary representation of ${\bf Z}$ (i.e. change by a phase
under $\t\rightarrow\t +n$) labelled by a parameter $e^{2\pi i\vt}\in U(1)$.
This is the familiar phenomenon of vacuum angles or $\vt$-vacua in an
embryonic setting.

In the twisted sectors one finds, instead of
(\ref{11}), the wave functions
\be
\Psi^{\vt}(\t,\ep) = \sum_{n}\exp{\left(-2 \pi^{2}\ep (n-\vt)^{2}\right) }
\exp{ \left(2\pi i(n-\vt) \t\right)} \label{var}
\ee
with the characteristic property
\[\Psi^{\vt}(\t+m,\ep)=\exp{\left(2\pi im\vt\right)}
\Psi^{\vt}(\t,\ep) \;\;.\]

All our calculations could equally well be carried out in one of the
twisted sectors of the theory but, as nothing is gained by this, we shall
concentrate on the invariant ($\vt =0$) sector in the following.

\noindent \underline{The Sphere}

Write $S^{2}=D_{1}\cup_{S^{1}}(-D_{2})$ and decompose the
delta function appearing in (\ref{4}) as
\be
\d\big(\frac{1}{2\pi}\int_{S^{2}}\xi - k\big)=\int_{-\infty}^{+\infty} d\t
\d \big(\frac{1}{2\pi}\int_{D}\xi- \t\big) \d\big(\frac{1}{2\pi}\int_{D'}\xi
- (\t-k) \big)\;\;.
               \label{23}
\ee
The two delta functions give rise to $\Psi_{D_{1}}(\t , \ep_{1})$ and
$\Psi_{-D_{2}}(k-\t , \ep_{2})$
respectively (cf. (\ref{10})), so that the partition function
$Z_{S^{2}}(k,\ep)$
(equation (\ref{5})) can be obtained from the wave functions on the disc by
\be
Z_{S^{2}}(k ,\ep) =\int_{-\infty}^{+\infty}d\t \Psi_{D_{1}}(\t ,
\ep_{1}) \Psi_{-D_{2}}(k-\t , \ep_{2})
           \label{24}
\ee
as can of course also be checked explicitly, with the help of the
Poisson summation formula
\be
\sum_{n} e^{-4\pi^{2}n^{2}t} \, e^{2\pi i n\t} = (4\pi t)^{-\frac{1}{2}}
\sum_{n} e^{-(\t +n)^{2}/4t} \, . \label{pois}
\ee

One may wonder what the calculation of $\int_{0}^{1}d\t \Psi_{D_{1}}^{P}(\t ,
\ep_{1}) \Psi_{-D_{2}}^{P}(k-\t , \ep_{2})$ results in. Note that
$\Psi^{P}(k-\t)=\Psi^{P}(-\t)=\Psi^{P}(\t)$, so that the difference among
the topological sectors is washed out in $\Psi^{P}$ and not unexpectedly
one then finds that
\be
\int_{0}^{1}d\t \Psi_{D_{1}}^{P}(\t ,
\ep_{1}) \Psi_{-D_{2}}^{P}(k-\t , \ep_{2}) =  \sum_{k} Z_{S^{2}}(k,\ep)
\;\;. \label{25}
\ee
Thus if one is only interested in results summed over all topological sectors
(as is frequently the case) $\Psi^{P}$ is adequate, but to get a handle
on the individual sectors we need to use $\Psi$.

\noindent \underline{Kernels On $\S_{g,n}$}

Denote by $\S_{g,n}$ a genus $g$ Riemann surface with $n$ boundary
components. Also we denote the partition function on such a manifold by
$K_{\S_{g,n}}$ if we use a standard delta function and by
$K_{\S_{g,n}}^{P}$ when the periodic delta function is used. The symbol
$Z_{\Sg}$ is reserved for the partition function of closed surfaces
(boundaryless).

{}From the derivation of (\ref{10}) and (\ref{11}), it is clear that
they are valid not only for the disc but more generally for a disc
with an arbitrary number of handles, i.e. for a surface $\Sigma_{g,1}$
surface. We thus have the general result
\bea
K_{\S_{g,1}}(\t, \ep)&=& \Psi(\t, \ep)\;\;,
                               \nonumber \\
K_{\S_{g,1}}^{P}(\t, \ep)&=& \Psi^{P}(\t, \ep)\;\;.\label{20}
\eea

The generalization to surfaces $\Sigma_{g,n}$ with $n>1$ boundaries is also
straightforward. In that case we have to specify $n$ holonomies
$\t_{1},\ldots,\t_{n}$. In the path integral, for a manifold with
$n$ boundaries,  when we change variables from
the gauge field to the field strength we find that
\be
\int_{\Sigma_{g,n}}\xi = \sum_{i=1}^{n} \oint_{\gamma_{i}} A \, ,
\label{hcon}
\ee
so that one must still integrate over $(n-1)$ gauge fields at the boundaries,
the $n$th being determined by the above relationship. We want to perform
the path integral
\be
\int DA D\f \, e^{L} \,
\prod_{i=1}^{n} \d\big(\frac{1}{2\pi}\oint_{\g_{i}}A -\t_{i}\big) \, .
\label{pih}
\ee

We may use the trivialising map to pass to the variable $\xi$ but this
still leaves the holonomies (\ref{pih}) to account for. We may interpret
this in the following way. On the manifold $\Sigma_{g,n}$, the gauge
invariant degrees of freedom of the gauge field are represented by the
holonomies and the field strength, subject to the one condition
(\ref{hcon}). On using the trivialising map, the path integral measure goes
over to
\be
\int DA \rightarrow \int D\xi \, \prod_{i=1}^{n}
D\left(\frac{1}{2\pi}\oint_{\g_{i}}A\right)  \; \,
\d\left(\frac{1}{2\pi}\int_{D}\xi - \frac{1}{2\pi}
\sum_{i=1}^{n}\oint_{\g_{i}}A
 \right) \, .
\ee
Integrating over the holonomies on the boundaries in (\ref{pih}) leaves us
with
\be
\int D\xi \, e^{L}\, \d\big(\frac{1}{2\pi}\int_{\Sigma_{g,n}}\xi - (\t_{1}+
\ldots +\t_{n }) \big) \, . \label{21}
\ee
This is easily done and one finds
\bea
K_{\Sigma_{g,n}}(\t_{1},\ldots,\t_{n}, \ep) &=& \Psi(\t_{1} + \dots
+ \t_{n}, \ep) \nonumber\\
 K_{\Sigma_{g,n}}^{P}(\t_{1},\ldots,\t_{n}, \ep) &=& \Psi^{P}(\t_{1} + \dots
+ \t_{n}, \ep)\label{22}\;\;.
\eea

\noindent \underline{Wilson Loops}

In order to calculate the expectation value of a contractible Wilson
loop $\exp{(i\a \oint_{\g}A)}$ on a surface $\S_{g,n}$, denoted by
\be
K_{\S_{g,n}}(\t_{1},\dots,\t_{n},\ep;\a) \, ,
\ee
we need only know what the expectation
value of the Wilson loop on the boundary of a disc is. Let the
expectation value of the Wilson loop on the disc be denoted by
\be
\Psi(\t,\ep;\a) \, .
\ee
Then evidently
\be
K_{\S_{g,n}}(\t_{1},\dots,\t_{n},\ep;\a) = \int_{-\infty}^{+\infty}
d\t_{n+1} K_{\S_{g,n+1}}(\t_{1},\dots,\t_{n},\t_{n+1} ,\ep_{1}) \Psi(
\t_{n+1} ,\ep_{2};\a) \, ,
\ee
with similar formulae in the case of the periodic kernels. It remains
only to determine $\Psi(\t,\ep;\a)$. But this requires no calculation,
for the boundary data of the disc path integral fixes $\oint_{\g}A =
2\pi \t$, so we have
\be
\Psi(\t,\ep;\a) = \exp{\left(2\pi i \a \t \right)}\Psi(\t,\ep) \, .
\ee

There are two cases for homologically non-trivial loops of charge $m \in Z$
on a surface $\S_{g,n}$. The first has to do with such loops that can be
pulled `off'. In this case simply attach a Wilson loop to one of the
boundaries, then convolute with a cylinder on that boundary to move the
loop `inside'. These manipulations give the result
\be
 \int_{-\infty}^{+\infty}
d\t K_{\S_{g,n}, \g}(\t_{1},\dots,\t_{n-1},\t,\ep_{1})\exp{\left(2\pi i \a
\t \right)} K_{C}(-\t,\t_{n},\ep_{2}) \, .
\ee

The second case is when the non-trivial loop cannot be pulled out of the
surface. In this case begin with the surface $\S_{g-1,n+2}$ and put a
Wilson loop on one its boundaries, say the $n+1$'th. The kernel for this is
\be
K^{P}_{\S_{g-1,n+2},\g}(\t_{1},\dots,\t_{n+2},\ep;m) =
\exp{\left(2\pi i m \t_{n+1}
\right)}K^{P}_{\S_{g,n+2}}(\t_{1},\dots,\t_{n+2}, \ep) \, .
\ee
The result we are looking for is obtained by convoluting the $n+1$
boundary component with the $n+2$, which lowers the boundary components
by $2$ but raises the genus by $1$ and at the same time introduces a
non-contractible Wilson loop into the surface. We get
\bea
& & \int_{0}^{1}
d\t K^{P}_{\S_{g-1,n+2}, \g}(\t_{1},\dots,\t_{n},\t, -\t,\ep;m) \nonumber \\
& & = \int_{0}^{1} d\t \exp{\left(2\pi i m \t_{n+1}
\right)}K^{P}_{\S_{g,n+2}}(\t_{1},\dots,\t_{n},\t,-\t, \ep) \nonumber \\
&= & \int_{0}^{1} d\t \exp{\left(2\pi i m \t_{n+1}
\right)}K^{P}_{\S_{g,n}}(\t_{1},\dots,\t_{n}, \ep) \nonumber \\
& & = \d_{m,0}K^{P}_{\S_{g,n}}(\t_{1},\dots,\t_{n}, \ep)
 \, .
\eea
This generalises the result that, for closed manifolds, non-trivial Wilson
loops have trivial expectation values (\ref{twl}).

\noindent \underline{Wilson Points}

It is clear that the expectation value of Wilson points on an arbitrary
surface is obtained by convoluting surfaces with more boundaries with
discs that have the Wilson points in them. So for us the expectation
value of some Wilson points on the disc is adequate. The calculation is
exactly the same as for the closed surfaces in the previous section. We
get for $n$ such points with charges $q_{i}$
\be
\Psi(\t+\sum_{i=1}^{n}q_{i},\ep)\, .
\ee
This result may be understood from the canonical quantization point of
view. $\f$ is the canonical conjugate momentum to $A$, so its action on
$\t$ is by differentiation, that is
\be
i\f \t = 2\pi \,  .
\ee
In terms of operators
the expectation value of the Wilson points on the disc takes the form
\bea
& & \exp{\left( \sum_{i=1}^{n} \frac{i}{2\pi}q_{i}\f(x_{i}) \right)} \Psi(\t,
\ep)
\nonumber \\
& & =
\exp{\left(  \sum_{i=1}^{n} q_{i}\frac{\del}{\del\t} \right)} \Psi(\t, \ep)
\nonumber \\
& & = \Psi(\t+\sum_{i=1}^{n}q_{i},\ep) \, .
\eea

\noindent\underline{Consistency Checks}

The next thing we check is the proper
behaviour of the kernels $K_{\S_{g,n}}$ under the operation of
glueing surfaces along boundaries. Again, in view of (\ref{20}) and (\ref{22}),
it is quite sufficient to check this in the particular case of two
cylinders $C_{1}$ and $C_{2}$ glued along a common boundary $\g_{1}$ to form a
cylinder $C$ with $A(C)=A(C_{1})+A(C_{2})$, or $\ep = \ep_{1}+ \ep_{2}$.
Writing
$\del C_{1} =\g_{0}+\g_{1},\;\del C_{2}=\g_{1}+\g_{2}$ one has
$\del C = \g_{0}+\g_{2}=\g_{0}+\g_{1}-\g_{1}+\g_{2}$, so that we expect
$K_{C}$ to be given by
\be
K_{C}(\t_{1},\t_{2}, \ep)=\int_{-\infty}^{+\infty}d\t
           K_{C_{1}}(\t_{1},\t, \ep_{1})K_{C_{2}}(-\t,\t_{2},\ep_{2}) \,
, \label{26}
\ee
and using (\ref{11}) and (\ref{22}) this can easily be verified explicitly.
Mutatis mutandis (\ref{26}) is valid for the glueing of any two surfaces to
form a surface with $n>0$ boundaries. It is also possible to consider the
joining of two boundaries of a surface,
$\Sigma_{g,n+2}\rightarrow\Sigma_{g+1,n}$. In general this is described by
a formula similar to (\ref{26}) (which we used in the calculation of
expectation values), but due to the linear and additive way in
which the holonomies enter into (\ref{22}) in the Abelian case, this simply
results in
\be
K_{\Sigma_{g+1,n}}(\t_{1},\ldots,\t_{n},\ep)=
              K_{\Sigma_{g,n+2}}(0,0,\t_{1},\ldots,\t_{n},\ep) \, , \label{27}
\ee
(and analogously for $K^{P}$).

Any other more complicated calculation can be reduced to a combination of
the three examples discussed above.

\setcounter{equation}{0}
\section{Partition Function In Yang-Mills Theory} \label{pfym}

For $U(1)$ gauge theory, we were able to evaluate the partition function
on an arbitrary genus surface directly. The global information about the
trivializing map was straightforward to encode. For trivial bundles,
this amounted to the observation that $\int_{\Sg}F_{A} = \int_{\Sg}dA=0
$, so that $\int_{\Sg}\xi =0$. This is a gauge invariant condition. For
$SU(n)$ bundles (or trivial $U(n)$ bundles) we would also expect a
condition of the form $\int_{\Sg} dA^{a}=0$ but this is clearly not
gauge invariant and it is far from obvious what one should take to be
the non-Abelian generalization of $\int_{\Sg}F_{A} =0$.

On manifolds with boundary, however, all the information that was
required of the trivialzing map, for the $U(1)$ theory, had to do with
gauge invariant boundary data. By glueing manifolds with boundary
together, it was possible to arrive at the results for the compact closed
manifolds. There is a direct non-Abelian generalization of this. Indeed
it is enough to know the result for the disc so as to generate the
results on arbitrary Riemann surfaces, with or without boundary. Recall
that identifying the sides of a cut Riemann surface gives back the
original Riemann surface. A cut Riemann surface is just a disc.

In this section we restrict our attention to Lie groups which are
compact, connected and simply connected.
%
All of the results obtained will be in terms of group representation
theory. The set of equivalence classes of irreducible unitary
representations of $G$ is denoted by $\G$. For $\l \in \G$, we denote by
$d(\l)$ the dimension of the representation, $\c_{\l}$ the character
(normalised by $\c_{\l}(1) =d(\l)$) and by $c_{2}(\l)$ the quadratic
Casimir invariant of $\l$. We use various properties of the characters
that are treated in detail in, for example, \cite{wal,tbtd}.

\noindent \underline{The Wave Function On The Disc}

Let $\g=\del D$ be the boundary of a disc $D$ ($\g \sim {\bf S}^{1}$).
Just as for the $U(1)$ theory, the only gauge invariant degree of
freedom of a gauge field on the circle is its holonomy. Choosing the
boundary condition to be $P \exp{(\oint_{\g}A)}=g_{1}\in G$ (modulo
conjugation, i.e. gauge transformations of $A$), our task is to compute
the path integral
\be
K_{D}(g_{1},A(D))\equiv\int_{\Phi}e^{L}\, \d(Pe^{\oint_{\g}A},g_{1}) \, .
\label{y5}
\ee

We need to specify the delta function that appears in (\ref{y5}) and
which is some delta function on the group $G$. There are two
possibilities. The first is to use the
delta function of $L^{2}(G)$, given in the spectral representation by
\be
\tilde{\d}(g,h)=\sum_{\l\in\G}d(\l)\c_{\l}(g^{-1}h)\label{y11}\;\;.
\ee
With this choice of delta function, the path integral (\ref{y5}) is not
manifestly conjugation invariant but, as the result turns out to be,
use of (\ref{y11}) is sufficient for our present purposes. We can however
build in conjugation invariance from the outset by using
the delta function $\d(g,h)$ on the space $L^{2}(G)^{G}$ of conjugation
invariant functions (class functions),
\be
\d(g,h)=\sum_{\l\in\G}\c_{\l}(g^{-1})\c_{\l}(h) \, , \label{y12}
\ee
related to $\tilde{\d}(g,h)$ by
\be
\d(g,h)=\int_{G}dg'\tilde{\d}(g,g'hg'^{-1}) \, , \label{y13}
\ee
as a consequence of the relation
\be
\int dg \c_{\l}(xgyg^{-1})=d(\l)^{-1}\c_{\l}(x)\c_{\l}(y)\;\;.\label{y14}
\ee
The group measure is normalised here so that the group volume is one
\be
\int_{G}dg = 1 \, .
\ee
Some consequences of changing this are explored later in this section.

In the case of surfaces with more than one boundary component, the use of
(\ref{y12}) actually becomes mandatory if one wants to work with the
gauge fixed path integral and retain conjugation invariance, as
explained in \cite{bt}. So here we will use the second alternative,
though for the disc both delta functions lead to the same results.

The boundary data is given in terms of the gauge potential. We need to specify
it in terms of the field strength. Using the non-Abelian Stokes' theorem,
this is possible in general on the disc and is explained in appendix
\ref{nonab}. The part we need is that the Schwinger-Fock gauge
\be
x^{\m}A_{\m}^{a}=0 \, ,
\ee
allows us to express the gauge field in terms of the field strength
\be
A_{\m}^{a}(x) = \int_{0}^{1} ds \, sx^{\n}F_{\n \m}^{a}(sx) \, .
\label{sfg}
\ee
The trivializing map is available, with $G^{a}(A)=x^{\m}A_{\m}^{a}$, so
that we obtain
\bea
& & K_{D}(g_{1},\ep) \nonumber \\
& & = \int D\xi \, D\f \, \exp{\left(\frac{1}{4\pi^{2}}
\int_{\Sg} i Tr \f \xi  + \frac{\ep}{8\pi^{2}}
\int_{\Sg } Tr \f * \f  \right)}
\sum_{\l\in\G}\c_{\l}(g^{-1})\c_{\l}(g_{1}) \, , \label{ymd}
\eea
with $g=P \exp{\left(\int_{\g}A\right)}$ and $A$ expressed in terms of
$\xi$ through (\ref{sfg}).

Let us now assemble the techniques that will go into computing (\ref{y5}).
It will be convenient to replace the path ordered exponential
in (\ref{y5}) by a quantum mechanics amplitude, namely
\bea
& & \c_{\l}\left(P \exp{\oint_{\g}A}\right) \nonumber \\
& & = \int D\e D\bar{\e}\, \exp{\left(i\int_{0}^{1}dt
[\bar{\eta}^{i}(t)\dot{\eta}^{i}(t) \right.} \nonumber \\
& & \left. \;\;\;\;\;\;\;\;-i
A_{\mu}^{a}(\g(t))\dot{\g}^{\mu}(t)\big(\bar{\e}\l^{a}
\e\big)(t)]\right) \bar{\e}^{i}(1)\e^{k}(0)\d_{ik} \, , \label{y6}
\eea
where $\e^{k}$ and $\bar{\e}^{k}$, $k=1,\ldots,d(\l)$ are Grassmann
variables (with the obvious generalization to traces of the form
$\c_{\l}(Pe^{\oint_{\g}A}g)$, $g\in G$) and where
$(\bar{\e}\l^{a}\e)(t)=  \bar{\e}^{i}(t)\l_{ik}^{a}\e^{k}(t)$.
A short proof of (\ref{y6}) uses the fact that the fermion propagator in
one dimension is
\[ \int D\e D\bar{\e}\exp{\left(i\int_{0}^{1}dt
\bar{\e}^{i}(t)\dot{\e}^{i}(t)\right)}
\bar{\e}^{i}(s) \e^{j}(0)=\delta^{ij} \theta(s)\;\;.\]
Together with the change of variables
\begin{eqnarray*}
\e_{i}(t) &\rightarrow &
\left[P\exp{(\int_{0}^{t}A_{\m}^{a}\l^{a}\dot{\g}^{\m}ds )}\right]_{ij}
\e^{j}(t) \, , \nonumber \\
\bar{\e}_{i}(t)&\rightarrow &
\bar{\e}^{j}(t)\left[P
\exp{(\int_{t}^{0}A_{\m}^{a}\l^{a}\dot{\g}^{\m}ds)}\right]_{ji} \, ,
\end{eqnarray*}
(path ordering is done from the lower end of the integral to the
upper regardless of which is greater) this can be seen to imply (\ref{y6}).

Using the Schwinger-Fock gauge allows us to write
\bea
& & \int_{0}^{1} dt A^{a}_{\m}(\g(t))\frac{\g^{\m}(t)}{dt}(\bar{\e} \l^{a}
\e) (t) \nonumber \\
& =&
\int_{0}^{1}dt \int_{0}^{1}ds \xi_{\n \m}^{a} s\g^{\n}
\frac{d\g(t)}{dt}(\bar{\e} \l^{a} \e)(t) \nonumber \\
&=& \int_{0}^{1}dt \int_{0}^{1}ds \xi_{\n \m}^{a} \frac{d(s\g^{\m})}{dt}
\frac{d(s\g^{\n})}{ds}(\bar{\e} \l^{a} \e)(t) \nonumber \\
&=& \int_{D} \xi^{a}  (\bar{\e} \l^{a} \e)(t) \, ,
\eea
with the local polar coordinates on the disc given by $s\g(t)$. To
evaluate (\ref{ymd}), we first perform the Gaussian integrals,
\bea
& & \int D\xi \, D\f \, \exp{\left(\frac{1}{4\pi^{2}}
\int_{D} i Tr \f \xi  + \frac{\ep}{8\pi^{2}}
\int_{D } Tr \f * \f  - i \int_{D} \xi^{a} (\bar{\e} \l^{a} \e)(t) \right)}
\nonumber \\
& & \; \; \; \; \; \; \; \; = \exp{\left( 2\pi^{2}\ep
\int_{0}^{1}dt(\bar{\e}\l^{a}\e)(t)(\bar{\e}\l^{a}\e)(t) \right)} \, ,
\eea
which then leaves us with the task of evaluating the fermionic integral
\[\int D\e D\bar{\e}\,
\exp{\left(i\int_{0}^{1}dt\bar{\e}^{k}(t)\dot{\e}^{k}(t)
+2\pi^{2}\ep \int_{0}^{1}dt(\bar{\e}\l^{a}\e)(t)(\bar{\e}
\l^{a}\e)(t) \right)}\, \,  \bar{\e}^{i}(1)\e^{k}(0) \, . \]
In a direct calculation for the cylinder \cite{bt}, one encounters a
slight generalization of this, namely
\bea
& & \int D\e D\bar{\e} \, \exp{\left(
\int_{0}^{1}dt[i\bar{\e}^{k}(t)\dot{\e}^{k}(t)
+\rho^{a}(t)(\bar{\e}\l^{a}\e)(t)] \right. } \nonumber \\
& &  \; \; \; \; \; \; \left.
+2\pi^{2}\ep \int_{0}^{1}dt(\bar{\e}\l^{a}\e)(t)(\bar{\e}\l^{a}\e)(t)
 \right) \, \bar{\e}^{i}(1)\e^{k}(0)\label{y15}\;.
\eea
These integrals can be calculated order by order in perturbation theory,
but in appendix \ref{piwl} we have given a simple derivation of the result
\be
(\ref{y15})= \exp{\left(-2\pi^{2}\ep c_{2}(\l)\right)}
[P\exp{\left(\int_{0}^{1}dt \rho(t) \right)}]^{ik} \;\;.
\label{y16}
\ee
Combining this with (\ref{y6}) and (\ref{y11}) or (\ref{y12}), we finally
arrive at the equation for the kernel (wave function) on the disc
(\ref{ymd}),
\be
K_{D}(g_{1}, \ep)=\sum_{\l\in\G}d(\l)\c_{\l}(g_{1}) \exp{\left(
-2\pi^{2} \ep c_{2}(\l) \right) } \label{y17}\;\;.
\ee

\noindent \underline{The Two Sphere}

$K_{D}$ can be used to compute the partition function of Yang-Mills theory
on $S^{2}$ as well as expectation values of Wilson loops. Considering $S^{2}$
as the union of two discs,
\[S^{2}=D_{1}\cup_{\g}D_{2}\;\;,\;\;\;\;\;\;\;\;\del D_{1} =\del (-D_{2})
=\g\;\;,\]
we see that we can write $Z_{S^{2}}$ as
\be
Z_{S^{2}}(\ep) =\int_{G}dg K_{D_{1}}(g,\ep_{1})K_{D_{2}}(g^{-1}, \ep_{2})
\, , \label{y20}
\ee
(the inverse $g^{-1}$ being due to the opposite orientation of $\del D_{2}$).
That $dg$, up to some overall constant, is the correct measure to use can
be seen from the change of variables $A\rightarrow P\exp{\int_{\g}A}$ on $\g$.
Using the orthonormality
\be
\int_{G}dg \c_{\l}(g)\overline{\c_{\m}(g)}=\int_{G}dg \c_{\l}(g)\c_{\m}(g^{-1})
=\d_{\l,\m}\label{y21}
\ee
of the characters this becomes
\be
Z_{S^{2}}(\ep)=\sum_{\l\in\G}d(\l)^{2}\exp{\left( -2\pi^{2}\ep
c_{2}(\l) \right)}\label{y22}\;\;,
\ee
where $\ep = \ep_{1} + \ep_{2}$ (this is the statement that $A(S^{2})
=  A(D)+A(D')$).

We are now in a position to determine the correct constraint on the field
$\xi$ that is needed so as to define the trivializing map directly on
the sphere. The required constraint can be deduced from inserting the
definition (\ref{y5}) of the kernel on the disc into (\ref{y20}). Doing
this, one finds (the connection $A$ should again be thought of as being
expressed in terms of $\xi$ via (\ref{eq:nicmap}))
\be
Z_{S^{2}}(\ep) =\int D\xi \exp{\left(+\frac{1}{8 \pi^{2} \ep }\int_{S^{2}}Tr
\xi*\xi \right)}
         \d\left(P\exp{\big(-\oint_{\del D}A\big)}, P\exp{\big(-\oint_{\del
D'}A\big)}\right)\label{y29}\;\; .
\ee
This constraint expresses the requirement that the holonomies of $A$
along $\del D_{1}$ and $\del(-D_{2})$ are equal up to conjugation.

We should like to express this in a form which makes transparent what
the condition on $\xi$ is. Via the non-Abelian Stokes'
theorem (details may be found in appendix \ref{nonab}),
the path ordered exponential entering (\ref{y29}) can be written as
\be
P\exp{(-\oint_{\del D_{1}}A)}={\cal P}\exp{(-\int_{D_{1}}\xi)} \, .\label{y30}
\ee
We therefore obtain
\be
Z_{S^{2}}(\ep) =\int D\xi e^{\left(+\frac{1}{8\pi^{2}\ep}\int_{S^{2}}Tr
\xi*\xi \right)}
         \d\left({\cal P}\exp{\big(-\int_{D_{1}}\xi\big)}{\cal
P}\exp{\big(-\int_{D_{2}}\xi\big)}\right) \label{y31}\;\;.
\ee
The splitting of $S^{2}$ into $D_{1}$ and $D_{2}$ is arbitrary here and for any
other choice of disc and complement this formula remains correct.

\noindent \underline{The Cylinder}

With this example we come to the heart of the matter. It is possible to
quite straightforwardly, following closely the analysis for the disc,
derive from scratch the partition function for the cylinder $C$ \cite{bt}.
However, such a direct approach is difficult to implement in the case of
higher genus surfaces, or for surfaces with more boundary components. For
that reason we will now give an evaluation of the partition
function on the cylinder which is based on nothing but the kernel for the
disc (\ref{y17}) and the fact that $K_{C}$ can depend only on the
holonomies along the boundaries and the area $A(C)$. These
considerations will be seen to generalize directly to any surface.

We deform the disc to a rectangle with the same area
with edges $a,b,c$ and $d$, that is, we view it as the cut surface of the
torus or of the cylinder. This is as in figure $2$, where $c$ is the
$a^{-1}$ cycle and $d$ is taken to be the $b^{-1}$ cycle. Write the
holonomy $g_{1}$ around the boundary
of $D$ as $g_{1}=g_{a}g_{b}g_{c}g_{d}$ (this is possible as
the holonomy is a path ordered exponential and can therefore be written as
the product of the group elements obtained from going along $a$, then along
$b$, etc.). Identifying the edges $a$ and $c$ (with opposite
orientation) now amounts to setting $g_{c}=g_{a}^{-1}$ and
integrating. Figures $5a$ and $5b$ give two ways of visualising this.
Using (\ref{y14}) and (\ref{y17}), we find
\bea
K_{C}(g_{b},g_{d}, \ep)
&=&\int_{G}dg_{a}K_{D}(g_{a}g_{b}g_{a}^{-1}g_{d}, \ep)
\nonumber\\
&=& \int_{G}dg_{a}\sum_{\l\in\G}d(\l)\c_{\l}(g_{a}g_{b}g_{a}^{-1}g_{d})
    \exp{\left(-2\pi^{2}\ep c_{2}(\l)\right) }\nonumber\\
&=& \sum_{\l\in\G}\c_{\l}(g_{b}) \c_{\l}(g_{d}) \exp{\left(
-2\pi^{2}\ep c_{2}(\l) \right) }    \label{y37}\;\; ,
\eea
which agrees with the more simple minded approach in \cite{bt}. We
should emphasise that this proceedure works, as Yang-Mills theory in two
dimensions is invariant under area preserving deformations.

There are a number of checks that can be made on this result. One we
mention here has to do with the axiomatic approach to topological field
theory. It is always possible to think of a genus $g$ Riemann surface as
a genus $g_{1}$ disc glued to one end of a cylinder and a genus $g_{2}$
disc glued at the other end, with $g=g_{1}+g_{2}$. Think of the discs
as generating states in the physical Hilbert space. The cylinder then has
the interpretation of an inner product between the `incoming' genus
$g_{1}$ state and the `outgoing' genus $g_{2}$ state. In the topological
limit $\ep \rightarrow 0$, (\ref{y37}) becomes
\be
\sum_{\l\in\G}\c_{\l}(g_{b}) \c_{\l}(g_{d}) \, ,
\ee
which is what we would expect. This simply says that the holonomies on
the left and right discs have to match up to conjugation. If we Fourier
transform, this is clear
\bea
C_{n m} &=& \int_{G}dg_{b} \int_{G} dg_{d} \bar{\c}_{n}(g_{b})
\bar{\c}_{m}(g_{d}) \sum_{\l\in\G}\c_{\l}(g_{b}) \c_{\l}(g_{d})
\nonumber \\
&=& \d_{n m} \, .
\eea

Another direct check of the method is to glue a disc to the $d$ end
of the cylinder, which yields a disc, and to see if this reproduces the
kernel for the disc (\ref{y17}). This indeed occurs
\bea
& & \int_{G} dg_{d} \sum_{\m} \c_{\m}(g_{d}^{-1})\exp{ \left(
-2\pi^{2}\ep_{1} c_{2}(\m) \right)
}  \sum_{\l\in\G}\c_{\l}(g_{b}) \c_{\l}(g_{d})
\exp{ \left( -2\pi^{2}\ep_{2} c_{2}(\l) \right) } \nonumber \\
& & \; \; \; \; = \sum_{\l\in\G} \c_{\l}(g_{b})
\exp{ \left( -2\pi^{2}\ep c_{2}(\l) \right) } \, ,
\eea
with $\ep = \ep_{1} + \ep_{2}$. Other tests may be found in \cite{bt}.

\noindent \underline{The Pants}

Does the same method allow us to calculate the kernel for the `pair of
pants' $\S_{0,3}$? Indeed it does. In figure $6$ we have exhibited one
possible cut
Riemann surface of the pants. Once more express the holonomy around the
boundary of the disc as the product of the holomies of the eight edges.
 From the figure we see that it is enough, once we visualize the cylinder
as a rectangle (disc) with a hole as in figure $5$, to identify the
marked edges, call them $a$ and $a^{-1}$, as above. Thus, to obtain
$K_{\S_{0,3 }}$, all we have to do
is calculate
\be
\int dg_{a}K_{C}(g_{a}g_{1}g_{a}^{-1}g_{2},g_{3}, \ep) \, ,
\ee
and, using (\ref{y37}) and (\ref{y14}), this becomes
\be
K_{\S_{0,3}}(g_{1},g_{2},g_{3}, \ep)
         =\sum_{\l\in\G}d(\l)^{-1}
         \c_{\l}(g_{1}) \c_{\l}(g_{2})
\c_{\l}(g_{3})\exp{ \left(-2\pi^{2}\ep c_{2}(\l) \right) }
\label{y39} \;\;.
\ee
Again one may check that glueing a disc to any of the ends reproduces
the kernel for the cylinder.

\noindent \underline{Extension To ${\bf \S_{g,n}}$}

Knowing the kernel of the `pants' and the rules for joining boundaries
and glueing surfaces it is now
a simple matter to deduce from (\ref{y39}) the general formulae
\be
K_{\S_{g,n}}(g_{1},\ldots,g_{n}, \ep)=
\sum_{\l\in\G}d(\l)^{2-2g-n}\c_{\l}(g_{1})\ldots \c_{\l}(g_{n}) \exp{
\left( -2\pi^{2}\ep c_{2}(\l) \right) } \, , \label{y40}
\ee
and
\be
Z_{\Sg}(\ep) =\sum_{\l\in\G}d(\l)^{2-2g}\exp{
\left( -2\pi^{2}\ep c_{2}(\l) \right) } \, . \label{y41}
\ee

It is rather remarkable that, in a sense, the basic building block of
Yang-Mills theory in two dimensions is not the kernel (\ref{y39})
of the `pants' but rather that of the disc (\ref{y17}). This
can be understood as a consequence of the
fact that the theory is not only almost topological in the above sense
but also a gauge theory.

Note that in (\ref{y40},\ref{y41}) the power
of $d(\l)$ is always the Euler number $2-2g-n$ of $\S_{g,n}$. That it is
precisely this function of $g$ and $n$ which appears is of course no
coincidence. Compatibility of (\ref{y40},\ref{y41}) with the operations of
joining $2b$ boundaries of a surface $\S_{g,n}$,
\[\S_{g,n}\rightarrow\S_{g+b,n-2b}\;\;,\]
and of glueing two surfaces $\S_{g,n}$ amd $\S_{g',n'}$  along $b$ boundaries,
\[(\S_{g,n},\S_{g',n'})\rightarrow \S_{g+g'+b-1,n+n'-2b}\;\;,\]
requires the putative power $p(g,n)$ of $d(\l)$ to satisfy
\bea
p(g,n)&=&p(g+b,n-2b)\nonumber\\
p(g,n)+p(g',n')&=& p(g+g'+b-1,n+n'-2b)
\eea
and this fixes $p(g,n)$ uniquely (up to a scale) to be
$p(g,n)=2-2g-n$. The scale can then be determined by computing e.g. the
kernel on the disc (\ref{y17}) or the partition function of the two-sphere
(\ref{y22}).

\noindent \underline{$vol \,{\cal M}\big(\Sg , SU(2)\big) $}

We specialise to the case that the structure group is $SU(2)$. Setting $\ep
= 0$, in the partition function (\ref{y41}), we get
\be
vol \, {\cal M}\big(\Sg , SU(2)\big)  \sim \sum_{\l\in SU(2)}
d(\l)^{2-2g} \,  .
\ee
The irreducible representations of $SU(2)$ are labeled by the positive
integers $n$ and the dimension of the $n$'th unitary irreducible
representation is $n+1$. There is a simple formula for
\be
\sum_{n=0}^{\infty} (n+1)^{2-2g}  \, ,
\ee
obtained by passing to the Riemann zeta function
\be
\zeta(s) = \sum_{n=1}^{\infty} n^{-s} \, , \; \; \; \; Re(s) >0 \, .
\ee
Thus we are interested in $\zeta(2g-2)$, which is related to the
Bernoulli polynomial $B_{2g-2}$ by
\be
\zeta(2g-2) = \frac{(2\pi)^{2g-2}}{2(2g-2)!} \mid B_{2g-2} \mid \, ,
\ee
our tentative expression for the volume being
\bea
vol \, {\cal M}\big(\Sg , SU(2)\big)  &\sim& \zeta(2g-2)
\nonumber \\
&=& \frac{(2\pi)^{2g-2}}{
2(2g-2 )!} \mid B_{2g-2} \mid \, . \label{vmod1}
\eea

This result is quite good. The volume of the moduli space is known, for
example by making use of Verlinde's formula, and is
\be
vol\, {\cal M}\big(\Sg , SU(2)\big)  = 2
\frac{\zeta(2g-2)}{(2\pi^{2 })^{g-1}} \, . \label{vmod2}
\ee
The factor of $2$ discrepency between (\ref{vmod1}) and (\ref{vmod2}) is
accounted for by noting that the centre of $SU(2)$ is $Z_{2}$ which has
order $2$, this being one of the factors we had previously mentioned but
omitted to carry around. The factor $(2\pi^{2})^{g-1}$ has a partial
explanation in terms of our inability to fix the normalisation of the group
integral used in the glueing rule. The exponent $g-1$ is determined in
this way but why it is $2\pi^{2}$ and not some other constant is difficult
to ascertain.

\noindent \underline{Standard Renormalizations}

There are two obvious source of arbitrariness in our calculations, thus
far, which we would now like to control. They have the same source,
namely, that we are not sure of the normalisation of the path integral
measure. The first is that the wavefunction or kernel on the disc should
be multiplied by an arbitrary constant $\kappa$. The second is that we
are also unable to fix the correct group measure in our glueing rules,
so let it be $\rho$ times the one thus far used.

We derive some consistency rules. If we glue two discs together, $D_{1}$
and $D_{2}$, along
one common edge to reproduce a new disc $D$, then the convolution of the
kernels
on the two discs should give the kernel on the new disc. This is the
case for the kernel (\ref{y17}) with group volume normalised to unity.
With the new normalisations we would have,
\be
\kappa_{D} K_{D}(\ep)  = \kappa_{D_{1}} \kappa_{D_{2}} \rho K_{D}(\ep) \,
,
\ee
which serves to fix the dependence of $\kappa$ and $\rho$ on the
parameters that are in the theory. If $\kappa$ is area dependent, then it
must be exponential, so set $\kappa_{D} = \exp{(v + bA(D))}$. Consistency is
achieved if $\rho = \exp{-v}$. If we demand the scaling symmetry that
relates the coupling constant to the area, then $bA(D)$ should be
replaced by $u\ep$ for some $u$. The net effect of this factor is to
multiply all of the previously derived kernels and partition functions
by $\exp{u \ep}$. In the topological limit this term plays no role.

We would like to work out the dependence on $\exp{v}$ for arbitrary
kernels. The way we do this is to begin with the kernel on a surface of
genus zero with $n$ boundaries, glue on a disc and demand it yields the
kernel on the zero genus surface with $n-1$ boundaries. It is not difficult
to see that all of these kernels are given by (\ref{y40}) with $g=0$
multiplied by $\exp{v}$. Higher genus surfaces are obtained in the
normal manner. Indeed for a closed manifold the result depends on the
genus only and is $\exp{v(2-2g)}$ times our previous result (\ref{y41}).

The ability to redefine the theory, by the introduction of the parameters
$u$ and $v$, may be viewed as the normal ambiguity one faces in using
different regularizations in any field theory. Changing the values of
$u$ and $v$ amounts to renormalization and Witten has dubbed these
variations, `standard renormalizations'. The factor $(2\pi^{2})^{1-g}$,
for example, may be obtained on setting $v= \frac{1}{2}ln(2 \pi^{2})$.
These considerations show, that if we know the volume of the moduli space
of flat connections for one surface (with $g \geq 2$), then all the
factors may be fixed. The Torus will not do, as $\exp{v(2-2g)} = 1$,
while for the sphere we run into $\zeta(-2)=0$.

\subsection{Expectation values of Observables}

For the case of the cohomological theories, with non-Abelian groups, we
saw that it is enough to know the partition function in order to
evaluate the observables of interest. So here we concentrate on the
observables that are intrinsic to the $BF$ theory, namely Wilson loops
and Wilson points. The situation is quite unlike the Abelian case as
here the Wilson loops are most certainly not trivial.

With the general formula (\ref{y40}) for $K(\S_{g,n})$ and the rules for
glueing surfaces and joining boundaries at our disposal, it is rather
straightforward to compute expectation values of Wilson loops (the
generalization to correlation functions of several non-intersecting
loops being immediate). There are three different types of
non-selfintersecting loops to consider, contractible (homotopically
trivial) loops, non-contractible homologically trivial loops, and
homologically non-trivial loops. As it is really homology and not homotopy
that matters, the first is actually a special case of the second type,
but for simplicity we will treat them separately.

\noindent\underline{Contractible Loops}

Expectation values of contractible loops on a surface $\S_{g}$ can be
computed by glueing a disc $\S_{0,1}$ and a $\S_{g,1}$ with a Wilson
loop on the boundary. We do the calculation for a Wilson loop on the
two-sphere. We want to compute the expectation value
\be
\langle\c_{\m}\left( P\exp{(\oint_{\g}A)} \right) \rangle_{S^{2}}\, .
\ee
We split $S^{2}$ along $\g$ into two discs $D_{1}$ and $D_{2}$ and put a
Wilson loop on the boundary of $D_{1}$ before glueing $D_{1}$ and $D_{2}$
together again. In equations this amounts to computing
\be
\langle\c_{\m}\left( P\exp{(\oint_{\g}A)} \right) \rangle_{S^{2},
\ep}=\int_{G}dg
 K_{D_{1}}(g, \ep_{1})\c_{\m}(g)K_{D_{2}}(g^{-1}, \ep_{2})\label{y23}\;\;.
\ee
In order to calculate this we make use of one further property of characters,
namely that
\be
\c_{\l}(g)\c_{\m}(g)=\c_{\l\ot\m}(g)\equiv\sum_{\rho\in\l\ot\m,\rho\in\G}
\c_{\rho}(g)\label{y24}\;\;.
\ee
Then we find
\be
\langle\c_{\m}\left( P\exp{(\oint_{\g}A)} \right) \rangle_{S^{2}, \ep}=
\sum_{\l\in\G}\sum_{\rho\in\l\ot\m}d(\l)d(\rho) \exp{\left(-2\pi^{2} \ep_{1}
c_{2}(\l)  \, -2\pi^{2} \ep_{2} c_{2}(\rho)  \right)}\label{y25}
\ee
for the unnormalized expectation value of a Wilson loop on $S^{2}$.

To get the result on a general surface one replaces $K_{D_{2}}$ in
(\ref{y23}) by $K_{\S_{g,1}}(\ep)$. Using the multiplicative property
(\ref{y24}) of characters we thus find
\bea
& & \langle\c_{\m}\left( P\exp{(\oint_{\g}A)} \right) \rangle_{\S_{g}, \ep}
\nonumber \\
& & = \int_{G}dg
 K_{D}(g,\ep_{D})\c_{\m}(g)K_{\S_{g,1}}(g^{-1}, \ep_{\S_{g,1}} )
\nonumber \\
& &=\sum_{\l\in\G}\sum_{\rho\in\l\ot\mu}d(\l)d(\rho)^{1-2g}
  \exp{\left(-2\pi^{2}(\ep_{D}
c_{2}(\l) +\ep_{\S_{g,1}}c_{2}(\rho) )  \right)} \, . \label{y42}
\eea
In the topological limit this is
\be
\sum_{\l\in\G}\sum_{\rho\in\l\ot\mu}d(\l)d(\rho)^{1-2g} \, ,
\ee
which because of the flatness condition should be $d(\m)$ times the
partition function.

\noindent\underline{Non-contractible Homologically Trivial Loops}

These types of loops exist on surfaces of genus $>1$ and cut a surface
$\S_{g'+g}$ into a $\S_{g',1}$ and a $\S_{g,1}$. Thus the only difference
to the example above is that we have to replace $D$ in (\ref{y42}) by
$\S_{g',1}$. This gives the result
\bea
& & \langle\c_{\m}(Pe^{\oint_{\g}A})\rangle_{\S_{g+g'}} \nonumber \\
& & \; \; =\sum_{\l\in\G}\sum_{\rho\in\l\ot\mu}d(\l)^{1-2g'}d(\rho)^{1-2g}
  \exp{\left(-2\pi^{2}(
 \ep_{\S_{g',1}} c_{2}(\l) + \ep_{\S_{g,1}} c_{2}(\rho) ) \right)}
 \, , \label{y43}
\eea
which reduces to (\ref{y42}) for $g'=0$.

\noindent\underline{Homologically Non-Trivial Loops}

Not unexpectedly the formulae in this case turn out to be slightly more
complicated than (\ref{y42},\ref{y43}). The required operation is now
not that of glueing two surfaces together but rather that of joining the
two boundaries of a $\S_{g-1,2}$ with a loop in between. In equations
this amounts to calculating
\bea
& & \langle\c_{\m}\left(P\exp{(\oint_{\g}A)} \right) \rangle_{\S_{g}, \ep}
\nonumber \\
& & = \int_{G}dg
 \sum_{\l\in\G}d(\l)^{2-2g}\c_{\l}(g)\c_{\m}(g)\c_{\l}(g^{-1})\exp{ \left(-
2\pi^{2} \ep c_{2}(\l) \right)}
 \nonumber\\
& & = \sum_{\l\in\G} \sum_{\rho\in\l\ot\mu} d(\l)^{2-2g} \d_{\l\rho}
\exp{\left(-2\pi^{2} \ep c_{2}(\l) \right)}
\;\;.\label{y44}
\eea
This means that a representation $\l\in\G$ will only contribute to the sum
if it appears again in the decomposition of $\l\ot\m$. Let $m_{\m}(\l)$
denote the multiplicity of $\l$ in $\l\ot\m$. Then
\be
\langle\c_{\m}\left(P\exp{(\oint_{\g}A)} \right) \rangle_{\S_{g}, \ep}=
     \sum_{\l\in\G}d(\l)^{2-2g}m_{\m}(\l) \exp{\left(- 2\pi^{2} \ep
c_{2}(\l)  \right)} \;\;.\label{y45}
\ee
For $SU(2)$, two
extreme cases are represented by choosing $\m$ to be a half-integer spin
representation or the spin one representation. If $\m$ is half-integer,
then for no value of $\l$ will $\l$ reappear in $\l\ot\m$, so that we
have the general result that for a homologically non-trivial loop $\g$
\[\langle\c_{n+\frac{1}{2}}\left( P\exp{(\oint_{\g}A)} \right)
\rangle_{\Sg, \ep}=0\;\;.\]
On the other hand if $\m=1$, then $m_{\m}(\l)=1 \;\forall\l\in\G$ and thus
all representations will contribute to the sum in (\ref{y45}),
\bea
\langle\c_{\m=1}\left( P\exp{(\oint_{\g}A)} \right) \rangle_{\Sg, \ep} &=&
               \sum_{\l\in\G}d(\l)^{2-2g} \exp{\left(-
2\pi^{2} \ep c_{2}(\l) \right)} \nonumber \\
&=& Z_{\Sg}(\ep) \;\;.
\eea
In the topological limit this gives back the volume of the moduli space.

The results of this section can of course also be used to calculate
correlation functions of several non-intersecting Wilson loops. The
intersecting case is more difficult but can be
dealt with at the level of the fermionic path integral
representation (\ref{y6}).

\noindent \underline{Wilson Points}

Let us work directly in the topological limit. In this case the position
of the Wilson points makes no difference to the result, so on the genus
$g$ Riemann surface we may as well consider all Wilson points to lie in a
disc. We proceed in the by now familiar fashion. We calculate the insertion
of the Wilson points on the disc and then we glue this to $\S_{g,1}$ to recover
the result on $\Sg$. We content ourselves with one insertion. The
general case is a simple extension.

We wish to calculate
\be
\int D\xi \, D\f \, \exp{\left(\frac{1}{4\pi^{2}}
\int_{D} i \f \xi  \right)} Tr_{R}\exp{(iq \f)} \,
\sum_{\l \in \G} \c_{\l}(g_{1}) \c_{\l}(P\exp{(\oint A)}) \, ,
\ee
on the disc and to do it we use the fermionic representations of the Wilson
points (\ref{wp}). The integral over $\f$ restricts $\xi$ to satisfy
(\ref{aflat}), which tells us that $\c_{\l}(P\exp{\oint A})$ does not
depend on the loop. The path integral reduces to
\bea
& & <0 \mid \bar{\e}^{i}\sum_{\l \in \G} \c_{\l}(g_{1}) \c_{\l}(\exp{q \e
.R.\bar{\e}} ) \e_{i} \mid 0> \nonumber \\
& & = \sum_{\l \in \G} \c_{\l}(g_{1})  Tr_{R} Tr_{\l} \exp{\left(q
R^{a}\otimes \l^{a}\right)} \, ,
\eea
and I leave it to the reader to disentangle this.

\setcounter{equation}{0}

\section{Instantons on complex K\"{a}hler surfaces}
There are natural generalizations of the two dimensional theories that
we have considered. These involve the moduli spaces of
Einstein-Hermitian structures \cite{kob} and of semi-stable holomorphic
bundles \cite{dk}. Here we will content ourselves with
a brief application to the moduli space of instantons over four
dimensional K\"{a}hler manifolds (complex K\"{a}hler surfaces).

Any two form $\Phi$ on an orientable four manifold may be decomposed into its
self-dual and anti-self-dual pieces
\be
\Phi^{+} = \frac{1}{2}(1+*) \Phi \; , \; \; \; \; \; \Phi^{-} =
\frac{1}{2}(1-*)
\Phi \, ,
\ee
by virtue of the fact that $*^{2}=1$ so that $\frac{1}{2}(1 \pm *)$ are
projection operators. Thus the space of two forms $\Omega^{2}(M)$
decomposes as
\be
\Omega^{2}(M) = \Omega^{2}_{+}(M) \oplus \Omega^{2}_{-}(M) \, .
\ee
Extending this to Lie algebra valued forms we have the decomposition
\be
\Omega^{2}(M,Lie G) = \Omega^{2}_{+}(M, Lie G) \oplus \Omega^{2}_{-}(M,
Lie G)\, .
\ee
In terms of this decomposition, the curvature tensor $F_{A}$ of a
connection on a bundle over $M$ may likewise be split and a connection
is said to be anti-self-dual (ASD) if
\be
F_{A}^{+}=0 \, .
\ee
An ASD connection is an anti-instanton.

On a complex manifold there is a second decomposition of $\Omega^{2}(M,Lie
G)$ available,
\be
\Omega^{2}(M,Lie G) = \Omega^{(2,0)}(M,Lie G) \oplus \Omega^{(1,1)}(M,Lie G)
\oplus \Omega^{(0,2)}(M,Lie G) \, .
\ee
The grading $(i,j)$ refers to the holomorphic and anti-holomorphic degrees so
that, for example,
\bea
& & F_{z_{1}z_{2}} dz_{1}dz_{2} \in \Omega^{(2,0)}(M,Lie G) \, , \nonumber
\\
& & F_{\bar{z}_{1}\bar{z}_{2}} d\bar{z}_{1}d\bar{z}_{2} \in
\Omega^{(0,2)}(M,Lie G) \, ,
\nonumber \\
& & F_{z_{i}\bar{z}_{j}} dz_{i}d\bar{z}_{j} \in \Omega^{(1,1)}(M,Lie G) \,
{}.
\eea
Furthermore, if the manifold $M$ is K\"{a}hler then it comes complete
with a non-degenerate closed two form $\omega$ of type $(1,1)$ so that
there is a further refinement
\be
\Omega^{(1,1)}(M,Lie G) = \Omega^{(1,1)}_{0}(M,Lie G) \oplus
\Omega^{0}(M,Lie G) . \omega \, ,
\ee
where $\Omega^{(1,1)}_{0}(M,Lie G)$ consists of the $(1,1)$ two forms
that are pointwise orthogonal to $\omega$. $\Omega^{0}(M,Lie G). \omega
$ is meant to indicate the component of the two form along the
K\"{a}hler form times the K\"{a}hler form.

The complex decomposition and the duality decomposition of
$\Omega^{2}(M,Lie G)$ are related by
\bea
\Omega^{2}_{+}(M,Lie G) &=& \Omega^{(2,0)}(M,Lie G) \oplus \Omega^{0}(M,Lie
G). \omega \oplus \Omega^{(0,2)}(M,Lie G) \, , \nonumber \\
\Omega^{2}_{-}(M,Lie G) &=& \Omega^{(1,1)}_{0}(M,Lie G) \, .
\eea

\noindent \underline{Anti-Instanton Moduli Space}

The ASD connections using the Riemannian structure of the four manifold
are, as we have seen, compactly written in terms of one equation
\be
F_{A}^{+} = 0 \, .
\ee
In terms of the K\"{a}hler structure of the manifold, the ASD
connections are specified by three equations, the first two being
\be
F^{(2,0)}_{A}=0 \, , \; \; \; F^{(0,2)}_{A} = 0 \, , \label{kinst2}
\ee
while the third $(F_{A},\omega) =0$, is neatly written as
\be
F_{A}\omega = 0 = \frac{1}{2}(F_{A},\omega) \omega^{2} \, .
\ee
The space of connections that satisfy the equations (\ref{kinst2}) is
denoted $\A^{(1,1)}$.

\noindent \underline{Topological Field Theory For ASD Instantons}

As $F_{A}\omega$ is a four-form, we may integrate it over the manifold.
This suggests that we take as an action the obvious generalization of
the two dimensional action namely
\be
\frac{i}{4\pi^{2}}\int_{M} Tr\left(  \f F_{A} \o + \frac{1}{2} \p
\p \o \right) + \frac{\ep}{8\pi^{2}}\int_{M} Tr \f ^{2}
\o ^{2} \, , \label{asd1}
\ee
where we have already included a term to take care of possible $\phi$ zero
modes. In order to impose the other two conditions (\ref{kinst2}), we
need to introduce two more Grassmann even fields $B^{(2,0)} \in
\Omega^{(2,0)}(M ,Lie G)$, $B^{(0,2)} \in \Omega^{(0,2)}(M,Lie G)$ and
two Grassmann odd fields $\chi^{(2,0)} \in \Omega^{(2,0)}(M ,Lie G)$,
$\chi^{(0,2)} \in \Omega^{(0,2)}(M,Lie G)  $. These are given the following
transformation rules
\bea
 & & \d \chi^{(2,0)} = B^{(2,0)} \, , \; \; \; \d B^{(2,0)} = [\chi^{(2,0)},
\phi] \, , \nonumber \\
 & & \d \chi^{(0,2)} = B^{(0,2)} \, , \; \; \; \d B^{(0,2)} =
[\chi^{(0,2)},\phi] \, ,
\eea
so that $\d$ continues to enjoy the property $\d ^{2} \Phi = {\cal
L}_{\phi} \Phi$.

We add to the action the following $\d$ exact term
\bea
 & &  \d \int_{M} Tr \left( \chi^{(2,0)} F^{(0,2)}_{A} + \chi^{(0,2)}
F^{(2,0)}_{A}
\right) \nonumber \\
 & & \; \; \; \; \; = \int_{M} Tr \left( B^{(2,0)} F^{(0,2)}_{A} + B^{(0,2)}
F^{(2,0)}_{A} \right. \nonumber \\
  & &  \; \; \; \; \; \; \; \; \; \left. - \chi^{(2,0)} (d_{A} \psi)^{(0,2)} +
\chi^{(0,2)} (d_{A}
\psi )^{(2,0)} \right) \, . \label{asd2}
\eea
Integration over the fields $B^{(2,0)}$ and $B^{(0,2)}$ forces the gauge
fields to satisfy (\ref{kinst2}) so that the path integral over $\A$ is
restricted to $\A ^{(1,1)}$. Likewise integration over the $\chi$ forces
the $\p$ to be tangents to $\A^{(1,1)}$.

The path integral that we have is then an interesting mixture of two
types of topological field theories. These correspond to the two types
of fixed point theorems that are available. The first part of the action
(\ref{asd1}), just as in the two dimensional theories, is a
Duistermaat-Heckman type action. The second term (\ref{asd2}) is of the
Matthai-Quillen form.

The analogy with the two dimensional version may be pushed further. The
critical points of (\ref{asd1}), taking into account (\ref{kinst2}) are
Yang-Mills connections. We can see this by noting that on ${\cal
A}^{(1,1)}$, $F_{A}\o$ is essentially $F_{A}^{+}$, so that on eliminating
$\f$ we produce an action which is, up to an additive constant, the
Yang-Mills action.

One more point worthy of note is the relationship between the
theory presented here and Donaldson theory (in K\"{a}hler form). These
are related to each other in the same way as the old and new versions of the
cohomological field theory in two dimensions as described at the end of
section \ref{coho}.

\noindent \underline{Observables}

For the case at hand, we may generalise the descent equation
(\ref{desc}) to
\be
(d + \d ) Tr (F_{A} + \p +\f)^{n} \omega^{m} = 0 \, , \label{desc4}
\ee
as $\omega$ is annihilated by both $d$ and $\d$. The topological observables
are (products of) integrals over some cycles in $M$ of $Tr (F_{A} + \p
+\f)^{n} \omega^{m} $, with the form of appropriate degree picked out,
which we write as
\be
\int_{\gamma} Tr (F_{A} + \p +\f)^{n} \omega^{m} \, . \label{obs4}
\ee

But what does
`topological' in this case refer to?  The theory described by
the combined action, (\ref{asd1}) and (\ref{asd2}), has an explicit
dependence on the K\"{a}hler form $\omega$ as do the observables
(\ref{obs4}). However, in this setting, the theory should depend not on
$\omega$ but rather on its cohomological class $[\omega] \in
H^{(1,1)}(M)$. This means that $\omega$ and $\omega + dK$, with $ dK \in
\Omega^{(1,1)}(M)$, should lead to the same results for the `topological'
observables. The difference of (\ref{obs4}) evaluated with $\omega$ and
evaluated with $\omega +dK$ is of the form
\be
\int_{\gamma} Tr (F_{A} + \p +\f)^{n} d X \, ,
\ee
for some $X$. Up to a sign this is
\be
\d \int_{\gamma} Tr (F_{A} + \p +\f)^{n} X \, ,
\ee
so that the difference is BRST exact and vanishes in the path
integral\footnote{ Notice that this derivation needs only that $dK \in
\Omega^{2}(M)$.}.
This derivation goes half way to showing that the action only picks up
$\d$ exact pieces, as we vary $\omega$ in its class, for (\ref{asd1}) is
exactly a combination of terms of the type (\ref{obs4}). As (\ref{asd2})
is in any case $\d$ exact we are done. This establishes
that the invariants will depend only on $[\omega]$.

There is the related issue of the dependence of the invariants and of
the action on the complex structure of $M$. An analysis of this issue is
possible along lines similar to
that of the dependence on the (almost) complex structure for the action
of the topological sigma models \cite{wsigma}. I will forgo this here.

\underline{Observables And The Partition Function}

The simplest observable is the path integral, with action the sum of
(\ref{asd1}) and (\ref{asd2}). For simplicity, consider the case where
$H^{2}_{A} =0$, that is where there are no zero modes at all of $B$,
$\chi$ or $\f$. The $B^{(2,0)}$ and $\chi^{(2,0)}$ integrals give us
\be
\d (F^{(2,0)}_{A}) \d( (d_{A} \p)^{(2,0)})  \, ,
\ee
which, off the zero set and around a prefered connection $A_{0} \in
\A ^{(1,1)}$, $A = A_{0} + a$, becomes
\be
\d ((d_{A_{0}} a)^{(2,0)}) \d ((d_{A_{0}} \p)^{(2,0)}) \, = \d(a^{(2,0)})
\d(\p^{(2,0)})
\, .
\ee
The determinants will exactly cancel (at the end of the day), up to a sign
which is not indicated.
The sign, however, is irrelevant, for when we take into account the $(0,2)$
contributions we will obtain the same sign which squares to unity. The path
integral is now over the (co-)tangent bundle of $\A ^{(1,1)}$, with the
action given by (\ref{asd1}).

We may set $\ep= 0$ with impunity and we do so. The
partition function is then equal to the symplectic volume of the moduli
space. This establishes that the simplest of Donaldsons invariants is
not zero (indeed is positive).

In the remainder we relax slightly the condition that $H^{2}_{A} = 0$ and
allow for the ``obstruction'' space $H^{2}_{A}$ to be made up entirely
of $H^{0}_{A} \omega$. In other words, we allow for $\f$ zero modes
(reducible connections) but not for $B$ or $\chi$ zero modes. The $B$
and $\chi$ fields may be integrated out as before and the partition
function of interest is
\be
Z_{M}(\ep) = \int_{T\A ^{(1,1)}} D\f \exp{\left( \frac{i}{4\pi^{2}}\int_{M}
 Tr \big(
\f  F \omega + \frac{1}{2} \p
\p \omega \big) +  \frac{\ep}{8\pi^{2}}\int_{M} Tr \f^{2}
\o ^{2} \right)} \, .
\ee

We do not evaluate this partition function, but rather can express other
observables in terms of it. The easiest examples are the expectation
values of products of ${\cal O}_{0}$. One may follow line for line the
steps in (\ref{ob1}) and (\ref{ob1a}) to obtain a formula in terms of
the differentiation of the partition function with respect to $\ep$.

If we could follow the steps of the two dimensional theory to calculate
the partition function, we would be able to go a long way towards
evaluating many of the Donaldson invariants of these moduli spaces.
Unfortunately our technology at the moment seems to be not up to this
task. The boundaries of four-manifolds being three-manifolds makes the
specification of the boundary data rather more involved. In this context
the work of Donaldson on the boundary value problem for Yang-Mills
fields may be helpful \cite{don}.

\appendix
\setcounter{equation}{0}
\section{Conventions} \label{note}

\noindent \underline{Lie Algebra Valued Fields}

When, in the text, a field $\f$ is said to be Lie algebra valued this
means
\be
\f = \f^{a}T_{a} \, ,
\ee
where the (anti-hermitian) $T_{a}$ are generators of the Lie algebra.
Commutators are Lie algebra brackets,
\be
[T_{a},T_{b}] = f^{c}\;_{ab}T_{c} \, ,
\ee
so that
\be
d_{A}\l = d\l + [A,\l] = (d\l^{a} + f^{a}\;_{bc}A^{b}\l^{c}) T_{a} \, ,
\ee
and
\be
F_{A}=dA+\frac{1}{2}[A,A] = (dA^{a} + \frac{1}{2} f^{a}\;_{bc} A^{b}
A^{c}) T_{a} \, .
\ee

\noindent \underline{Local Coordinate Expressions}

In the text differential form notation has been used. For those who
prefer explicit index notation, we give the correspondences here.

A zero-form is a function. Any one-form $A$ has as a local expression
\be
A = A_{\m}dx^{\m} \, , \nonumber
\ee
while a two form $B$ is
\be
B=B_{\m \n} \, dx^{\m} \, dx^{\n} \,. \label{2f}
\ee
The differentials $dx^{\m}$ anti-commute amongst themselves, so that
only the antisymmetric part of $B_{\m \n}$ appears in (\ref{2f}). The
differential $d$ is
\be
d = dx^{\m} \partial_{\m} \, ,
\ee
and squares to zero. With these rules we have
\bea
F_{A}& =& dA + \frac{1}{2}[A,A] \nonumber \\
&=& (\partial_{\m}A_{\n} +\frac{1}{2} [A_{\m},A_{\n}]) dx^{\m}dx^{\n}
\nonumber \\
&=&\frac{1}{2} F_{\m \n} dx^{\m}dx^{\n} \, .
\eea
The local gauge transformation, for the gauge field, becomes
\be
\d_{\l}A = d\l + [A,\l] = D_{\m}\l dx^{\m}\, ,
\ee
with
\be
D_{\m}= \partial_{\m} + [A_{\m}, \; \; \; \; \; .
\ee

Given a metric $g_{\m \n}$ on the manifold we also have the Hodge $*$
operator that in $n$ dimensions maps
$p$-forms to $(n-p)$-forms. Its action is defined by
\be
*(dx^{\m_{1}}\dots dx^{\m_{p}}) = \frac{\sqrt{\det{g}}}{(n-p)!}
\ep^{\m_{1} \dots \m_{p}}\;_{\m_{p+1} \dots \m_{n}}
dx^{\m_{p+1}} \dots
dx^{\m_{n}} \, ,
\ee
where $\det{g} \equiv \det{g_{\m \n}} $ and the epsilon symbol with all
the labels down is the antisymmetric matrix density with entries $(0,1,-1)$
when
any labels are repeated, or they are in an even permutation, or an odd
permutation, respectively. For example, in two dimensions
$\ep_{11}=\ep_{22}=0$ and $\ep_{12}=-\ep_{21}=1$. One raises the labels
with the metric tensor, so that
\be
\ep_{\m_{1} \dots \m_{n}} =  \det{g} \; \ep^{\m_{1} \dots \m_{p}} \, .
\ee
The invariant volume element is $\sqrt{\det{g}}dx^{\m_{1}}\dots
dx^{\m_{n}}$ which is often written as $\sqrt{\det{g}}d^{n}x$.

The following are now easily derived
\bea
\int_{\Sg} \f * \f& = &\int_{\Sg} \sqrt{\det{g}} d^{2}x \f^{2}(x) \,
\nonumber \\
\int_{\Sg} \f F_{A}&=& \int_{\Sg} d^{2}x \f F_{1 2} \, .
\eea
We also have, in two dimensions
\be
*d_{A_{0}}*A_{q} = \nabla _{\mu} A^{\mu} \, ,
\ee
where $\nabla _{\mu}$ is the covariant derivative in the metric sense,
and also covariant with respect to $A_{0}$, while the Yang-Mills
equations read
\be
*d_{A}*F_{A} = \frac{1}{2 }\nabla _{\mu}F^{\mu} \; _{\n} dx^{\n} \, .
\ee

\noindent \underline{Instantons And The Symplectic Form}

Let us fix on $R^{4}$ the standard coordinates $x^{\m}$, and on the
complex $2$-plane the coordinates $z^{1} = x^{1}+ix^{2}$, $z^{2} = x^{3}
+i x^{4}$. The $(2,0)$ and $(0,2)$ forms are spanned by
\bea
dz_{1}dz_{2}& =& (dx^{1}dx^{3}-dx^{2}dx^{4}) +i
(dx^{1}dx^{4}+dx^{2}dx^{3}) \, , \nonumber \\
d\bar{z}_{1} d\bar{z}_{2} &=& (dx^{1}dx^{3}-dx^{2}dx^{4}) -i
(dx^{1}dx^{4}+dx^{2}dx^{3}) \, .
\eea
The symplectic $2$-form is
\be
\o = \frac{i}{2}dz_{1}d\bar{z}_{1} + \frac{i}{2}dz_{2}d\bar{z}_{2} =
dx^{1}dx^{2} + dx^{3}dx^{4} \, .
\ee

The self-dual two forms $\Phi$ satisfy
\be
\Phi_{\a \beta} = \frac{1}{2} \ep_{\a \beta \g \d}\Phi^{\g \d} \, ,
\ee
so that
\bea
\Phi& =& 2\Phi_{1 2}(dx^{1}dx^{2} +dx^{3}dx^{4}) + 2
\Phi_{13}(dx^{1}dx^{3} -dx{2}dx^{4}) +
2\Phi_{14}(dx^{1}dx^{4}+dx^{2}dx^{3}) \nonumber \\
&=& i \Phi_{12}\o + (\Phi_{13} -i \Phi_{14})dz_{1}dz_{2} + (\Phi_{13} +i
\Phi_{14}) d\bar{z}_{1} d\bar{z}_{2} \, .
\eea
This is the decomposition advertised in the text.

\setcounter{equation}{0}
\section{Non-Abelian Stokes' Theorem} \label{nonab}

In the following we are working on a contractible manifold $M$ of dimension
$m$ or, equivalently, consider what follows to be performed on a single
coordinate neighbourhood,
which is diffeomorphic to an open set in {\bf  R}$^{m}$. Given
any gauge field (connection) $A$ on a principle $G$ bundle over $M$, there
is a gauge transformed connection $A^{U}= U^{-1}AU + U^{-1}dU$ such that
\be
x.A^{U}=0 \, . \label{eq:ap1}
\ee

This (Schwinger-Fock) gauge allows us to represent the gauge field $A$ in
terms of the field strength (curvature) $F_{A}$ and the group element $U$.
Equation (\ref{eq:ap1}) serves as a definition of $U$. It is in
terms of these quantities that the non-Abelian Stokes' theorem is
stated.

\noindent \underline{Abelian Stokes' Theorem}

The integral around a loop $\gamma$ of a connection $A$ is, by
Stokes'theorem, equated with the integral over any surface $\Gamma$ with
boundary $\partial \Gamma = \gamma$ of the curvature $F_{A}=dA$,
\be
\int_{\gamma}A = \int_{\Gamma}F_{A} \, .
\ee
Alternatively for Wilson loops this is
\be
\exp{(\int_{\gamma}A)} = \exp{(\int_{\Gamma}F_{A})} \, ,
\ee
and it is this formula that generalises, in a gauge invariant way, to the
non-Abelian case.

\noindent \underline{Non-Abelian Stokes' theorem}

Before turning to this let us make one observation. Within $M$ the surface
$\Gamma$ may be quite contorted. However by a suitable choice of local
coordinate functions it may be taken to be the unit disc in an {\bf
R}$^{2}$ plane of {\bf R}$^{m}$ centred at the origin. We work with these
local coordinates.

As it is not, perhaps, apparent that one may specify any connection in
terms of its curvature and a group element (corresponding to the usual
gauge freedom) we show this first. We express $A^{U}$ in terms of
$F_{A}^{U} = F(A^{U})=U^{-1}F_{A}U$,
\bea
A_{\mu}^{U} &=& \int_{0}^{1} ds \frac{d}{ds}[A_{\mu}^{U}(sx)s] \nonumber \\
&=& \int_{0}^{1} ds [sx^{\nu}\partial A_{\mu}^{U}(sx)/
\partial(sx^{\nu}) + A_{\mu}^{U}(sx)] \nonumber \\
&=& \int_{0}^{1} ds [sx^{\nu}\partial A_{\mu}^{U}(sx)/
\partial(sx^{\nu}) - sx^{\mu}\partial A_{\nu}^{U}(sx)/
\partial(sx^{\mu})] \nonumber \\
&=& \int_{0}^{1} ds sx^{\nu}F_{\nu \mu}^{U}(sx) \label{eq:ap2}
\eea
where
\be
F_{\nu \mu}^{U}(sx) = \frac{\partial}{\partial sx^{\n}} A^{U}_{\mu}(sx)
 - \frac{\partial}{\partial sx^{\m}}
A^{U}_{\nu}(sx)  + [A^{U}_{\nu}(sx),A^{U}_{\mu}(sx)] \,
.  \label{eq:ap3}
\ee
The third line in (\ref{eq:ap2}) follows by differentiation of
(\ref{eq:ap1}) at the point $sx$, that is, $A^{U}_{\mu}(sx) +
sx^{\nu} \partial A^{U}_{\nu}(sx)/ \partial (sx^{\mu}) =0$, while the last
line follows from the fact that $sx^{\mu} [A^{U}_{\mu}(sx) , A^{U}_{\nu}
(sx)] =0 $. In terms of the original field we have
\be
A_{\mu}(x) = U(x) \int_{0}^{1}ds\, sx^{\nu} U^{-1}(sx)F_{\nu \mu}(sx)U(sx)
U^{-1}(x) -  \partial_{\mu}U(x) U^{-1}(x) \, , \label{eq:ap4}
\ee
though this may be unedifying.

More interesting for us is the application of these ideas to the path
ordered exponential
\be
P\exp{(\oint_{\gamma_{x}} A)} \, , \label{eq:ap5}
\ee
around a (necessarily) contractible loop $\gamma$ with prefered point
$x$. The path ordered exponentials for $A$ and $A^{U}$ are related by
\be
P\exp{(\oint_{\gamma_{x}} A^{U})} = U^{-1}(x)P\exp{(\oint_{\gamma_{x}}
A)} U(x) \, ,  \label{eq:ap6}
\ee
or
\bea
& & P\exp{(\oint_{\gamma_{x}} A)} \nonumber \\
& & = U(x) P\exp{(\oint_{\gamma_{x}} A^{U})}
U^{-1}(x) \nonumber \\
& & = U(x) P\exp{(\oint_{\gamma_{x}} \int_{0}^{1} s\gamma^{\nu}U^{-1}(s\gamma)
F_{\nu \mu}(s\gamma) U(s\gamma) ds d\gamma^{\mu} )} U^{-1}(x) \, .
\label{eq:ap7}
\eea
The last equality is known as the non-Abelian Stokes' theorem. This
terminology is justified by noting that in the Abelian case this
equivalence reduces to the usual Stokes' theorem. Let us parameterise
the boundary curve (unit circle) by $t$. The local coordinates $x^{\mu}$
restricted to the disc are given in terms of $s$ and $t$ by $x^{\mu}= s
\gamma^{\mu}(t)$. The $s$ coordinate is `radial' while $t$ is `angular'. In
this way we see that for Abelian groups, where path ordering is
irrelevant (all the matrices commute so their order is immaterial), the
exponents appearing in (\ref{eq:ap7}) may be equated as
\bea
\int_{\gamma}A &=& \oint A_{\mu}(\gamma(t)) d\gamma(t) \nonumber \\
&=& \oint \int_{0}^{1}F_{\nu \mu}(s\gamma(t)) s\gamma^{\nu}(t)
\frac{d}{dt}\gamma^{\mu}(t) ds dt \nonumber \\
&=& \oint \int_{0}^{1}F_{\nu \mu}(s\gamma(t))
\frac{d}{dt}(s\gamma^{\mu}(t))\frac{d}{ds}(s\gamma^{\nu}(t)) ds dt
\nonumber  \\
&=& \int_{\Gamma} F_{A} \, .
\eea

The gauge invariant version of the non-Abelian Stokes' theorem is
obtained by taking the trace on both sides of (\ref{eq:ap7})
\bea
Tr P\exp{(\oint_{\gamma_{x}} A)}& =& Tr P\exp{(\oint_{\gamma_{x}} \int_{0}^{1}
 s\gamma^{\nu}U^{-1}(s\gamma)
F_{\nu \mu}(s\gamma) U(s\gamma) ds d\gamma^{\mu} )}  \nonumber \\
&=& Tr {\cal P} \exp{ \int_{\Gamma} F_{A}^{U}}\, ,
\label{eq:ap8}
\eea
with the second line defining what is meant by the surface ordered
exponential.

Alternative derivations may be found in \cite{ar,men}.

\setcounter{equation}{0}
\section{Laplacian on $G$ and the Schr\"{o}dinger equation on the disc}
\label{lap}
The Schr\"{o}dinger equation that we are interested in is
\be
[\frac{\partial}{\partial t} \, + \, \oint \frac{\delta}{\delta
A}.\frac{\delta}{\delta A} ] \Psi \, = \, 0 \, , \label{eq:b1}
\ee
where $\Psi$ is gauge invariant and depends on $A$ only through its
holonomy on the boundary of
the disc and $t$ represents some `evolution' from the centre of the
disc. We want to relate the solutions of (\ref{eq:b1}) to the
eigenfunctions of the Laplacian on the group $G$. These are the
characters of $G$,
\be
\Delta_{G} \chi_{\lambda}(g) = c_{2}(\lambda) \chi_{\lambda}(g) \, ,
\label{eq:b2}
\ee
where $c_{2}$ is the quadratic Casimir of the representaion.

By the Peter-Weyl theorem $\Psi$ must have the form
\be
\Psi(g) = \sum_{\lambda}a_{\lambda} \chi_{\lambda}(g) \, , \label{eq:b3}
\ee
where the $a_{\lambda}$ are functions of $t$ and
\be
g = P exp {\oint A} \, . \label{eq:b4}
\ee
The $A$ represent tangent space variables to the group elements, and, in
particular, for those group elements of the form (\ref{eq:b4}), the
Laplacian at $g$ is just the flat (tangent space) Laplacian so that
\be
\oint \frac{\delta}{\delta A}.\frac{\delta}{\delta A} \chi_{\lambda}(g)
=  c_{2}(\lambda) \chi_{\lambda}(g) \, . \label{eq:b5}
\ee
This may be obtained explicitly by noting
\be
\frac{\delta}{\delta A^{a }(x)} Tr P \exp{(\oint A)} = Tr P \lambda^{a}
\exp{(\oint_{x}A)} \, , \label{eq:b6}
\ee
with the notation $\oint_{x}$ indicating that the path begins at $x$.

The Schr\"{o}dinger equation becomes (by orthogonality of the characters)
\be
[\frac{\partial}{\partial t} + c_{2}(\lambda)] a_{\lambda}=0 \, ,
\ee
so that the most general solution is
\be
\Psi(t,g) = \sum_{\lambda}c_{\lambda} \chi_{\lambda}(g)
\exp{(-tc_{2}(\lambda))} \, ,
\ee
with the $c_{\lambda}$ constants.

\setcounter{equation}{0}
\section{Path integral Representation of Wilson loops} \label{piwl}
In the text we introduced a path integral representation for the
character of the holonomy of the gauge field in a given representation
$\lambda$ of the structure group. The solution of the following path
integral was needed at various points of the analysis

\bea
[\Psi_{\lambda}(\rho,\ep)]^{ij}& =& \int D\e D\bar{\e}
\exp{\left(\int_{0}^{1}dt [i\bar {\e}^{k}(t)\dot{\e}^{k}(t)
+\rho^{a}(t)(\bar{\e}\l^{a}\e)(t)] \right. } \nonumber \\
 &  & \; \; \left. - \frac{\ep}{2}
 \int_{0}^{1}dt (\bar{\e}\l^{a}\e)(t)(\bar{\e}\l^{a}\e)(t) \right)
\bar{\e}^{i}(1)\e^{k}(0) \label{eq:c1}\; .
\eea
Consider first the trace of this $\Psi_{\lambda}(\rho,\ep) =
\Psi_{\lambda}(\rho,\ep)^{ij}\d_{ij}$. We do not need to evaluate the
trace of (\ref{eq:c1}). Rather we note that it satisfies the
Schr\"{o}dinger equation (\ref{eq:b1}) with the initial condition that it is
the character
\be
\Psi_{\lambda}(\rho,0)= \chi_{\lambda}(P\exp{\oint \rho}) \, .
\ee
The solution, following our previous analysis, is thus
\be
\Psi_{\lambda}(\rho,\ep) = \chi_{\lambda}(P\exp{\oint \rho})
\exp{(-\frac{\ep}{2}c_{2}(\lambda))} \, .
\ee

To see that $\Psi_{\lambda}(\rho,\ep)$ satisfies (\ref{eq:b1})
(with $t= \ep/2$) firstly differentiate (\ref{eq:c1}) with respect to
$\ep/2$ to obtain
\bea
\frac{\partial}{\partial \ep/2} \Psi_{\lambda}(\rho,\ep)& =&
\langle \int_{0}^{1}dt (\bar{\e}\l^{a}\e)(t)(\bar{\e}\l^{a}\e)(t)
\rangle  \nonumber \\
&=&  \oint \frac{\d^{2}}{\d \rho^{2}}\Psi_{\lambda}(\rho,\ep)\, ,
\eea
where $\langle Z \rangle$ means the insertion of the field $Z$ in the
path integral on the right hand side of (\ref{eq:c1}).

When considering the matrix elements $\Psi_{\lambda}(\rho,\ep)^{ij}$, we
use the same argument with the initial condition that at $\ep=0$ this is
\be
\Psi_{\lambda}(\rho,0)^{ij}= [P \exp{\oint \rho}]^{ij} \, ,
\ee
to arrive at
\be
\Psi_{\lambda}(\rho,\ep)^{ij}= [P \exp{\oint \rho}]^{ij}
\exp{\left(-\frac{\ep}{2}c_{2}(\l) \right)} \, .
\ee

\underline{{\bf Figure Captions}}

\underline{Fig. 1}
A genus $g$ surface with a particular choice of the homology basis.

\underline{Fig. 2}
The homology basis for the Torus with its associated cut surface.

\underline{Fig. 3}
Homology basis for a genus $2$ surface.

\underline{Fig. 4}
The cut Riemann surface associated to the genus $2$ surface.

\underline{Fig. 5a,b}
Two ways of seeing the identification of edges of a disc to form a cylinder.

\underline{Fig. 6}
A possible identification of the edges of a disc to obtain the pants.


\begin{thebibliography}{99}
\bibitem{schwarz} A.S. Schwarz, {\em The partition function of a
               degenerate quadratic functional and Ray-Singer
               invariants}, Lett. Math. Phys. 2 (1978) 247.
\bibitem{rst}  D.B. Ray and I.M. Singer, {\em R-torsion and the Laplacian on
               Riemannian manifolds} . Math. 7 (1971) 145.
\bibitem{wtft} E. Witten, {\em Topological quantum field theory},
               Commun. Math. Phys. 117 (1988) 353.
\bibitem{d} S.K. Donaldson, {\em A polynomial invariant for smooth four
            manifolds}, Topology 29 (1990) 257.
\bibitem{wjp} E. Witten, {\em Quantum field theory and the Jones
            polynomial}, Commun. Math. Phys. 121 (1989) 351.
\bibitem{bt1}  M. Blau and G. Thompson, {\em Topological gauge theories of
               antisymmetric tensor fields}, Ann. Phys. 205 (1991) 130.
\bibitem{h} G.T. Horowitz, {\em Exactly soluble diffeomorphism invariant
               theories}, Commun. Math. Phys. 125 (1989) 417.
\bibitem{bt2}  M. Blau and G. Thompson, {\em A new class of topological field
               theories and the Ray-Singer torsion}, Phys. Lett. 228B (1989)
                64.
\bibitem{hs} G.T. Horowitz and M. Srednicki, {\em A quantum field
                theoretic description of linking numbers and their
               generalization}, Commun. Math. Phys. 130 (1990) 83.
\bibitem{wtpg} E. Witten, {\em On the structure of the topological phase
               of two dimensional gravity}, Nucl. Phys. B340 (1990) 281.
\bibitem{dijk} R. Dijkgraaf, {Intersection theory, integrable
              hierarchies and topological field theory}, preprint
              IASSNS-HEP-91/91.
\bibitem{as} M.F. Atiyah and I.M. Singer, {\em Dirac operators coupled
              to vector potentials}, Proc. Nat'l Acad. Sci. (USA) 81
              (1984) 2597.
\bibitem{nr} N.S. Narasimhan and T.R. Ramadas, {\em Geometry of $SU(2)$
           gauge fields}, Commun. Math. Phys. 67 (1979) 121, T.R. Ramadas,
             Thesis, Tata Institute, unpublished.
\bibitem{bv} O. Babelon and C.M. Viallet, {\em The Riemannian geometry
            of the configuration space of gauge theories}, Commun. Math. Phys.
            81 (1981) 515.
\bibitem{gp} D. Groisser and T.H. Parker, {\em The Riemannian geometry
             of the Yang-Mills moduli space}, Commun. Math. Phys. 112
            (1987) 663.
\bibitem{bs} L. Baulieu and I.M. Singer, {\em Topological Yang-Mills
             symmetry}, Nucl. Phys. Suppl. 5B (1988) 12.
\bibitem{brt} D. Birmingham, M. Rakowski and G. Thompson, {\em BRST
              quantization of topological field theories}, Nucl. Phys.
             B315 (1989) 577.
\bibitem{bbt} D. Birmingham, M. Blau and G. Thompson, {\em Geometry and
              quantization of topological gauge theories}, Int. J. Mod.
              Phys. A5 (1990) 4721.
\bibitem{ot} S. Ouvry and G. Thompson, {\em Topological field theory:
              zero modes and renormalization}, Nucl. Phys. B344 (1990)371.
\bibitem{kanno} H. Kanno, {\em Weyl algebra stucture and geometrical
             meaning of BRST transformation in topological quantum field
             theory }, Z. Phys. C43 (1989) 477.
\bibitem{bta} M. Blau and G. Thompson, {\em $N=2$ topological gauge
             theory, the Euler characteristic of moduli spaces and the
             Casson invariant}, preprint NIKHEF-H/91-28, MZ-TH/91-40, to
             appear in Commun. Math. Phys..
\bibitem{btb} M. Blau and G. Thompson, {\em Topological gauge
             theories from supersymmetric quantum mechanics on spaces of
             connections}, preprint MZ-TH/91-42, to appear in Int. J.
             Mod Phys. A.
\bibitem{bbrt} D. Birmingham, M. Blau, M. Rakowski and G. Thompson, {\em
               Topological field theory}, Phys. Reports 209 (1991) 129.
\bibitem{w2d} E. Witten, {\em On quantum gauge theories in two dimensions},
             Commun. Math. Phys. 141 (1991) 153.
\bibitem{w2dr} E. Witten, {\em Two dimensional gauge theories revisited},
             preprint IASSNS-HEP-92/.
\bibitem{ab} M.F. Atiyah and R. Bott, {\em The Yang-Mills equations over
             Riemann surfaces},
             Phil. Trans. R. Soc. Lond. A308 (1982) 523.
\bibitem{R} B. Rusakov, {\em Loop averages and partition functions in
            $U(N)$ gauge theory on two-dimensional manifolds}, Mod.
            Phys. Lett. A5 (1990) 693.
\bibitem{M} A.A. Migdal, {\em Recursion relations in gauge field
            theories}, Sov. Phys. JETP 42 (1976) 413.
\bibitem{wheater} J.F. Wheater, {\em Two dimensional lattice gauge theories
            are topological}, Oxford preprint OUTP-90-19P; {\em Topology
            and two dimensional lattice gauge theories}, Oxford preprint
            OUTP-91-07P.
\bibitem{bralic} N.E. Bralic, {\em Exact computation of loop averages in
               two-dimensional Yang-Mills theory}, Phys. Rev. D 22 (1980)
              3090.
\bibitem{raj} S.G. Rajeev, {\em Yang-Mills theory on a cylinder},
              Phys. Lett. 212B (1988) 203.
\bibitem{bt} M. Blau and G. Thompson, {\em Quantum Yang-Mills theory on
             arbitrary surfaces}, Int. J. Mod. Phys. A (1992) .
\bibitem{fine} D.S. Fine, {\em Quantum Yang-Mills on the two-sphere},
               Commun. Math. Phys. 134 (1990) 273, {\em Quantum
               Yang-Mills on a Riemann surface}, Commun. Math. Phys. 140
              (1991) 321.
\bibitem{deg} P. Degiovanni, {\em Th\'{e}ories topologiques,
              d\'{e}veloppments r\'{e}cents}, unpublished lecture notes
               (1991).
\bibitem{aj}  M.F. Atiyah and L. Jeffrey, {\em Topological Lagrangians
              and cohomology}, J. Geom. Phys. 7 (1990) 120.
\bibitem{mq} V. Mathai and D. Quillen, {\em Superconnections, Thom
             classes and equivariant differential forms}, Topology 25
            (1986) 85.
\bibitem{b} M. Blau, {\em The Mathai-Quillen formalism and topological
              field theory}, lecture notes at the Karpac winter school,
             preprint NIKHEF-H/92-07.
\bibitem{ab2} M.F. Atiyah and R. Bott, {\em The momentum map and
               equivariant cohomology}, Topology 23 (1984) 1.
\bibitem{dh} J.J. Duistermaat and G.J. Heckmann, {\em On the
               variation in the cohomology in the symplectic form of the
              reduced phase space}, Invent. Math. 69 (1982) 259.
\bibitem{btym2} M. Blau and G. Thompson, unpublished.
\bibitem{ms} D. Montano and J. Sonnenschein, {\em The topology of moduli
              space and quantum field theory}, Nucl. Phys. B313 (1989)
               258; J. Sonnenscein, {\em Topological quantum field
              theories, moduli spaces and flat connections}, Phys. Rev.
             D42 (1990) 2080.
\bibitem{vafa} C. Vafa, {\em Operator formulation on Riemann surfaces},
               Phys. Lett. 190B (1987) 47.
\bibitem{cs} J.M.F. Labastida and A.V. Ramallo, {\em Operator formalism for
             Chern-Simons theories} Phys. Lett. 227B (1989) 92.
\bibitem{at} M.F. Atiyah, {\em Topological quantum field theories},
            Publ. Math. I.H.E.S. 68 (1988) 175.
\bibitem{wal} N.R. Wallach, {\em Harmonic Analysis on Homogeneous Spaces},
              Marcel Dekker, New York (1973).
\bibitem{tbtd} T. Br\"{o}cker and T. tom Dieck, {\em Representations of
             compact Lie groups}, Springer Verlag, New York (1985).
\bibitem{kob} S. Kobayashi, {\em Differential Geometry of complex vector
            bundles }, Princeton University Press, (1987).
\bibitem{dk} S.K. Donaldson and P.B. Kronheimer, {\em The geometry of
             four-manifolds}, Oxford University Press, Oxford (1990).
\bibitem{wsigma} E. Witten, {\em Topological Sigma Models}, Commun.
              Math. Phys. 118 (1988) 411.
\bibitem{don} S.K. Donaldson, {\em Boundary value problems for
            Yang-Mills fields}, J. Geom and Phys. 8 (1992) 89.
\bibitem{ar} I.Ya. Aref'eva, {\em Non-Abelian Stokes formula}, Teor.
            Mat. Fiz. 43 (1980) 111.
\bibitem{men} M.B. Mensky, {\em Application of the group of paths to the
             gauge theory and quarks}, Lett. Math. Phys. 3 (1979) 513.


\end{thebibliography}
\end{document}